\documentclass[aps,prx,twocolumn,superscriptaddress]{revtex4-2}

\usepackage{graphicx}
\usepackage[font={small}]{caption}
\usepackage{subcaption}
\usepackage{amsmath}
\usepackage{amssymb}
\usepackage{tabularx}
\usepackage{placeins}
\usepackage{units}
\usepackage{multirow}
\usepackage{braket}
\usepackage{color}
\usepackage{hyperref}
\usepackage[normalem]{ulem}
\usepackage{breakcites}
\usepackage{adjustbox}
\usepackage{mathtools}
\usepackage{float}
\usepackage{adjustbox}
\usepackage{dcolumn}
\usepackage{bm}

\begin{document}

\title{A new generation of effective core potentials: selected lanthanides and heavy elements}

\author{Haihan Zhou$^1$ } 
\email{hzhou23@ncsu.edu}
\author{Benjamin Kincaid$^1$, Guangming Wang$^1$, Abdulgani Annaberdiyev$^2$, Panchapakesan Ganesh$^2$, and Lubos Mitas$^1$ }
\affiliation{
1) Department of Physics, North Carolina State University, Raleigh, North Carolina 27695-8202, USA \\
2) Center for Nanophase Materials Sciences Division, Oak Ridge National Laboratory, Oak Ridge, Tennessee 37831, USA \\
}

\begin{abstract}

We construct correlation-consistent effective core potentials (ccECPs) for a selected set of heavy atoms and $f$-elements that are of significant current interest in materials and chemical applications, including Y, Zr, Nb, Rh, Ta, Re, Pt, Gd, and Tb.
As customary, ccECPs consist of spin-orbit averaged relativistic effective potential (AREP) and effective spin-orbit (SO) terms. 
For the AREP part, our constructions are carried out within a relativistic coupled-cluster framework while also taking into objective function one-particle characteristics for improved convergence in optimizations.
The transferability is adjusted using binding curves of hydride and oxide molecules.
We address the difficulties encountered with $f$-elements, such as the presence of large cores and multiple near-degeneracies of excited levels.
For these elements, we construct ccECPs with core-valence partitioning that includes $4f$-subshell in the valence space.
The developed ccECPs achieve an excellent balance between accuracy, size of the valence space, and transferability and are also suitable to be used in plane wave codes with reasonable energy cutoffs.
\end{abstract}

\maketitle

\section{Introduction}
Several years ago, we introduced correlation-consistent effective core potentials (ccECPs) that are constructed in an explicit many-body framework in contrast to more traditional one-particle approaches \cite{Bennett2017, Bennett2018, Annaberdiyev2018, Wang2019, Wang2022}.
Our effort has been motivated by the use of valence-only calculations in both stochastic many-body methods such as quantum Monte Carlo (QMC) as well commonly used quantum chemistry approaches such as Configuration Interaction (CI) and Coupled Cluster (CC).
Besides many-body constructions with overall high accuracy, this new generation of effective valence-only Hamiltonians also offers testing and benchmarking on molecular systems to boost the transferability for real chemical and material applications. 

We have also established sets of reference data for constructed ccECPs, such as exact/nearly-exact total and kinetic energies and corresponding values typically obtained in single-reference real-space QMC calculations within the fixed-node approximation \cite{Annaberdiyev2020}.
Similarly, our high accuracy QMC calculations of solids provided another set of transferability checks \cite{Annaberdiyev2021, Annaberdiyev2022, Melton2020} in addition to other many-body calculations on large molecular complexes and solids carried out by independent groups \cite{Wang2021, Li2022, Wines2020, Oqmhula2020, Otis2020, Wang2019a, Zhou2019, Tyagi2023, Shin2021, Hill2022, Goldzak2022, Denawi2023, Huang2021}.
We also probed the limits of systematic fidelity to all-electron results for $3sp$ main group elements with large cores that showed biases due to the absence of shallow core states in the valence space \cite{Bennett2018}.
For example, significant biases were found for Al, Si, etc., with [Ne]-core in oxide molecules that show clear overbinding trends at smaller bond lengths.
To overcome this limitation, we have also introduced small-core options with [He] instead of [Ne] cores \cite{Bennett2018}.
Another aspect that has been  important in our effort is the issue of broad usability. For this purpose, accurate Gaussian basis sets have been provided. 
In addition, for plane-wave calculations with deeper semi-core states included in valence space, we offer another soft-ccECP version adapted for low cut-off in plane wave codes \cite{Kincaid2022}.

Here we present ccECPs for another selected set of elements, including transition metals, heavy elements, and, most importantly, a few lanthanides. 
It is well-known that the construction of ECPs for these elements is challenging due to several reasons, such as large relativistic effects including spin-orbit, many valence electrons, and also very large cores that can affect the transferability due to core polarizations and/or relaxations in varying chemical environments; difficult to resolve near-degeneracies showing up for states with occupations by $4f$, $5d$, $6s$ and $6p$ orbitals. 
Another issue is that there are much fewer calculations for heavy elements that probe the fidelity of ECPs in general, especially when compared with calculations involving elements from the first three rows \cite{Dolg1989, Dolg1989a, Dolg1993, Gibson2003, Huelsen2009, Lu2019, Lu2022}.
For example, recent $f$-element calculations by Bauschlicher \cite{Bauschlicher2022} show that some of the ECPs developed during the 1980s turned out to be less accurate than perhaps expected. One can see  that the calculations are very sensitive to the choice of basis and that the resulting data can be significantly biased. This provides a clear motivation for revisiting this part of the periodic table with new generation ECP construction.
In essence, our effort is focused on providing another alternative with better testing and benchmarking data from the outset to improve overall prospects of the constructed valence-only Hamiltonians. 
In this study, we opted for larger core sizes to clearly delineate the achievable accuracy of such a choice. 
Our well-tested construction can provide predictive power for many-body calculations across various applications regardless of minor compromises on the accuracy side.

In what follows, we briefly summarize the methods used. These are primarily based on our previous work \cite{Wang2022}. We then describe the resulting operators, check the discrepancies of binding curves for hydride and oxide molecules, and make some overall accuracy comparisons with a summary of the results.

\section{f-elements core choice}

Obvious complexities for heavier atoms arise from the increasing number of electrons in the core resulting from filling the shells with higher principal numbers. 
This leads to the associated core and semi-core polarizability and relaxation  effects, especially in the lanthanide row where $f$-electrons are involved.
Although most of the $f$-electron charges are in the core region, the tails extend far beyond the nominal core radii and typically reach  bonding regions \cite{Lu2019, Lu2022}.
Hence, for $4f$ elements, an appropriate choice of the core  size vs. valence space is less obvious. 
One possible choice is to decrease the core size to the innermost 28 electrons, i.e., all levels up to the principal quantum number $n=3$.
For such a choice of core the valence space will include several dozen electrons that leads to large total energies, which, unfortunately, makes it very costly to use methods such as QMC. Since core shells contributions scale as $\propto Z^2$ where $Z$ is the atomic number,  corresponding fluctuations decrease the QMC calculations,
see, for example \cite{Foulkes2001}.
We note that overall QMC scaling on number of electrons $N$ is very favorable
$\mathcal{O}(N^2{\;\rm to\;} N^3)$\cite{Foulkes2001},
depending on details of the calculation. We also note that efficient coupled cluster algorithms with single, double and perturbative triple excitations scale as $\mathcal{O}(N^6 {\;\rm to\;} N^7)$\cite{Lesiuk2020}. However, this favorable QMC scaling assumes a limited energy range (eg,  valence band(s) states in a solid), otherwise the cost increases rapidly. For example, standard implementation of all-electron diffusion QMC for an atom scales as $\approx Z^{5.5}$,
\cite{Foulkes2001}.

On the other hand, many constructions opt to use large cores, which also include $f$-electrons \cite{Lu2022, Huelsen2009, Dolg1989}.
Although the resulting number of valence electrons is then small, which might be adequate for certain applications, the absence of the $f$-shell, especially in the middle of the lanthanide series, is problematic for many applications.
In particular, lanthanides are some of the key elements for high-quality magnets where the large moments of the partially filled $f$-shell are crucial for properly describing magnetic properties.
Since hybridization often changes the $f$-shell occupation, including these electronic states is essentially unavoidable in most  materials calculations.


In our constructions, we attempt to balance accuracy and core size. Therefore, we have chosen to probe for a compromise solution that has been tested on the Tb element \cite{Annaberdiyev2023}, namely using $5s5p4f$ one-particle levels in the valence 
for elements with partially filled $f$-subshells. This choice was less explored previously; for example, it was not included for lanthanides
in the well-known Stuttgart-Koeln-Bonn energy consistent ECP table \cite{Leininger1997,Stoll2002}.
Our work tests the viability of such core-valence partitioning with encouraging outcomes.
In addition, we expect that this will be further ascertained in the future through more extensive calculations and tests.

\section{Methods}

\subsection{ECP form}
The construction of our ECPs aims to reproduce valence properties of the all-electron Hamiltonian for given elements in a many-body, correlated framework. The form of the pseudopotentials follows the same parameterization as used in our previous works \cite{Annaberdiyev2018}.
It consists of two parts, averaged relativistic effective potential (AREP) and corresponding spin-orbit (SO) terms:
\begin{equation}
    V^{SOREP} =  V^{AREP} + V^{SO}.
\end{equation}
This can be further explicitly written as: 
\begin{equation}
V^{AREP}(r) = V_{loc}(r) + \sum_{l = 0}^{\ell_{\text{max}}=L-1} (V_l(r) - V_{loc}(r))\sum_{m}\ket{lm}\bra{lm}
\end{equation}
Here the term $V_{loc}$ represents the local potential, while  $V_l - V_{loc}$ stands for the non-local $s$, $p$, $d$ and $f$ potentials with $\sum_m \ket{lm} \bra{lm}$ as the angular-momentum projector.
$\ell_{max}$ ranges from 3 to 4 depending on the element in question.
$V_{loc}$ is chosen to remove the Coulomb singularity from the potential.
The local contribution is given as follows: 
\begin{equation}
    V_{loc}(r) = - \dfrac{Z_{\rm eff}}{r} (1-e^{\alpha r^2}) + \alpha Z_{\rm eff} r e^{-\beta r^2} + \sum ^{2} _{i=1} {\gamma_i e ^{-\delta_i r^2}}
\end{equation}
where $Z_{\rm eff}$ is the number of electrons in the valence space. 
$\alpha$, $\beta$, $\delta_i$, and $\gamma_i$ are all parameters optimized during the optimization.
This form ensures the perfect cancellation of the Coulomb singularity and smoothes out the curvature of the potential at small $r$ \cite{Burkatzki2008}. 

The non-local channels follow a similar form and are parameterized as given by 
\begin{equation}
    V_l (r) = \sum _{j=1} {\beta_{\ell j} r ^{n_{\ell j} - 2} e ^{-\alpha_{\ell j} r^2}}
\end{equation}
where $n_{\ell j}$ are fixed integers determined prior to optimization, and $\alpha_{\ell j}$, $\beta_{\ell j}$ are parameters optimized for each non-local channel individually.
Most functions in each channel have $n_{\ell j}$ equal to 2.

The spin-orbit terms are derived from Lee's definition of the general SOREP\cite{Ermler1981, Lee2008} form. In our work, it is given by
\begin{equation}
V^{SO} = \sum_{l =1}^{l_{max}^{'}} \Delta V_{l} \mathcal{P}_{l} \Vec{l} \cdot \Vec{s} \mathcal{P}_{l}
\end{equation}
where $l_{max}^{'}$ is a chosen integer that indicates the number of channels of spin-orbit terms. $\mathcal{P}_l$ is the same angular-momentum projector described above. The radial potential in the spin-orbit terms can be further parameterized as:

\begin{equation}
\Delta V_l = \sum_{k} \beta_{lk} r^{n_{lk} - 2} e^{-\alpha_{lk} r^2}
\end{equation}

Note that in both AREP and SO term constructions, the parameters $n_{lj}$ and $n_{lk}$ are usually chosen as integers and fixed during the process. The set of parameter $\alpha_{lj}$, $\alpha_{lk}$, $\beta_{lj}$ and $\beta_{lk}$ forms our parameter space. DONLP2\cite{Spellucci2009}, a sequential quadratic programming (SQP) algorithm based optimizer, is used to find the best combination of parameters with respect to the objective functions we constructed. For all the elements, we decided to find the best parameters of the spin-orbit terms after constructing the AREP ECP.
As indicated in our previous work, the AREP part is most decisive regarding accuracy and transferability.

For Y, Zr, Nb, Rh, Ta, Re, and Pt, since the choice of their core is the same as the elements in the same row in our previous works, the objective functions and optimization process are similar. However, for Tb and Gd, the novel core choice also has led to extra complexities. To have a better starting point, two different initialization strategies have been experimented with before the legacy optimization workflow.
The construction of objective functions in the legacy optimization routine will be introduced in the next section; Tb and Gd will be discussed shortly after.

\subsection{Objective functions}

The objective functions are crucial in constructing ECPs, as they directly reflect the desired properties to be reproduced. For the AREP ECP optimizations, the objective functions consist of several items. One of the important components is the minimization of atomic spectra discrepancies with respect to the ground state  calculated at the CCSD(T) level. This includes the electron affinity (EA), several excited states, and ionization potentials (IPs) that aim to capture frequently observed effective charge states of the given atom in molecules and solids.  Another useful criterium 
is reproducing the single-particle eigenvalues since they determine the tails of one-particle states and indirectly also the charge norm conservation.
The objective function can be expressed as:
\begin{equation}
\mathcal{O}^2_{AREP} = \sum_{s \in S} \omega_{s} (\Delta E_{ECP}^s - \Delta E_{AE}^s)^2 + \sum_{i \in I} \omega_{i} (\epsilon_{ECP}^i -\epsilon_{AE}^i) 
\end{equation}

Here $\Delta E_{\cdot}^s$ represents the atomic gap of a state  $s$ with respect to the ground state in ECP and AE cases.
Hence the first term indicates the sum of squared biases of the chosen states from the AE atomic spectrum.
The corresponding weight $\omega_s$ can be tuned to adjust the influence of each particular state, and having this flexibility improves the convergence of optimizations. $\epsilon_{\cdot}^i$ represents the single-particle eigenvalue in ECP and AE cases. Usually, they correspond to the ground state, while $\{\omega_{i}\}$ are the corresponding weights. In our optimizations, $\{ \omega_i\}$ are typically larger than $\{ \omega_s\}$, with the difference reaching up to an order of magnitude to balance the impact of the terms that leads to faster optimization.

Here all the energies are obtained from CCSD(T) level calculations with large uncontracted basis sets, typically aug-cc-pwCVTZ with extra diffusive terms. For Gd and Tb, due to the sensitivity of the basis set choice, we have expanded the basis set to aug-cc-pwCVQZ with diffusive terms. All of our CCSD(T) calculations are carried out by Molpro v.2015.1 with 10$^{th}$ DKH Hamiltonian accounting for scalar-relativistic effects. 

Very often, our ECP parameters are initialized from Stuttgart ECPs and further trained with the target objective functions. In some of the cases, we also tried to start with parameters from previous ccECPs of other elements, for example, in the same group or close in atomic number. Using this approach, Zr, Nb, Rh, Ta, and Re ECPs have been successfully constructed.

The spin-orbit terms are subsequently optimized after the AREP ccECPs are optimized and verified with transferability tests. Similar to the above, the objective function for the spin-orbit terms can be written as:

\begin{align}
\mathcal{O}^2_{SO} = \sum_{s \in S} \omega_{s} (\Delta E_{ECP}^s - \Delta E_{AE}^s)^2 \nonumber \\ + \sum_{m \in M} \omega_{m} (\Delta E_{ECP}^m -\Delta E_{AE}^m) 
\end{align}

The main difference here is that instead of single-particle eigenvalues, we focus on the multiplet splitting energies referenced from the lowest energy of the same electronic state in ECP and AE cases, represented by $\Delta E_{\cdot}^{m}$. All the calculations for spin-orbit optimizations are carried out using the DIRAC code at the complete open-shell configuration interaction (COSCI) level of accuracy with the exact two-component Hamiltonian (X2C). 

After the AREP and SOREP constructions are finalized, well-tempered (aug)-cc-p(C)V$n$Z basis sets ($n$ = D, T, Q, 5 for non-$f$ elements and D, T, Q for $f$-elements) are generated.
The basis sets are optimized using the CISD method.
We optimize the lowest exponent and the ratio between two subsequent exponents for each basis channel.
After this procedure, the augmentation terms were added.
More detailed explanations of basis set generations are given in the Supplementary Material.

\subsection{ECP initialization of Tb and Gd}
\label{ECP_init_5f}

For the non-$f$ elements, the initial parameters are often obtained from Stuttgart ECPs or previous ccECPs for other elements.
However, due to the special choice of core for $f$-elements, namely [Kr]$4d^{10}$ (46 electrons), the initialization of ECP parameters becomes a non-trivial challenge.
In the first place, we generate all the initial parameters from scratch. Additionally, in the complicated space formed by ECP parameters, numerous local minima can easily trap the optimization process if the initial guesses are not close enough to a desired or, at least, an acceptable local minimum.

For the Tb element, the initial guess generation can be summarized as follows:
\begin{enumerate}
    \item Divide the collection of chosen states $S$ into subsets $S_i$ according to changing occupation in the channels with ${i \in {s, p, d, f}}$.
    \item For each subset $S_i$, use a reduced objective function $\mathcal{O}_{AREP}^i$ and optimize this single channel. 
    \item After all the non-local channels are independently optimized, combine all of them and initialize a local channel. Optimize the combined ccECP with the original objective function $\mathcal{O}_{AREP}$.
\end{enumerate}

The primary gain here is reducing the number of parameters and the fact that the objective function $\mathcal{O}_{AREP}^i$ contains only one single-particle eigenvalue. This makes the optimization process finish much faster than if there more parameters and a more complicated objective function. The results for the Tb element will be shown in later sections. 

However, it is not guaranteed that this will work for all cases of interest. 
This happens when the above procedure is applied to initialize the Gd element.
After many restarts, we tried a slightly modified approach, which eventually worked as follows:

\begin{enumerate}
    \item Partition the chosen set of states $S$ into $S_{sp}$, $S_{d}$ and $S_{f}$. 
    \item We optimize the $S_{sp}$ objective function with extra single-particle eigenvalues from different states included, beyond just the ground state.
    \item Add the $d$ channel with initial guesses and update the modified reduced objective function with $d$-related atomic spectrum gaps
    so that $spd$ channels are optimized.
    \item Update the $f$ channel and repeat step 3.
\end{enumerate}


These additional optimization steps for the $f$-elements lead to a unique set of difficulties during ECP constructions.
Currently, due to the complicated parameter space, we found that it is difficult to come up with a universal solution to the initialization problem of the $f$-elements.
At the same time, since Tb and Gd have been generated successfully, as will be shown later, these can be used as possible starting points for other $f$-elements.

\section{Results}\label{Results}
For each element we demonstrate the accuracy of the spectrum in a table with three versions of absolute deviations related to the ECP gaps compared to the corresponding AE values.
The first one is the mean absolute deviation (MAD). For the considered $N$ atomic states, the MAD is defined as follows:
\begin{equation}
    \mathrm{MAD}=\frac{1}{N} \sum_{s=1}^{N}\left|\Delta E_{s}^{\mathrm{ECP}}-\Delta E_{s}^{\mathrm{AE}}\right| .
\end{equation}
The next metric is the MAD of selected low-lying gaps for a smaller subset of $n$ atomic states that includes only the states of the first excitation state (EX, mostly $s \leftrightarrow d$ transitions), electron affinity (EA), first ionization potential (IP), and second ionization potential (IP2):
\begin{equation}
    \mathrm{LMAD}=\frac{1}{n} \sum_{s=1}^{n}\left|\Delta E_{s}^{\mathrm{ECP}}-\Delta E_{s}^{\mathrm{AE}}\right| .
\end{equation}
Additionally, a weighted MAD (WMAD) is defined as well for all considered N gaps as follows:
\begin{equation}
    \mathrm{WMAD}=\frac{1}{N} \sum_{s=1}^{N} \frac{100\%}{\sqrt{\left|\Delta E_{s}^{\mathrm{AE}}\right|}} \left|\Delta E_{s}^{\mathrm{ECP}}-\Delta E_{s}^{\mathrm{AE}}\right| .
\end{equation}

\begin{figure}[!htbp]
\centering
\includegraphics[width=1.00\columnwidth]{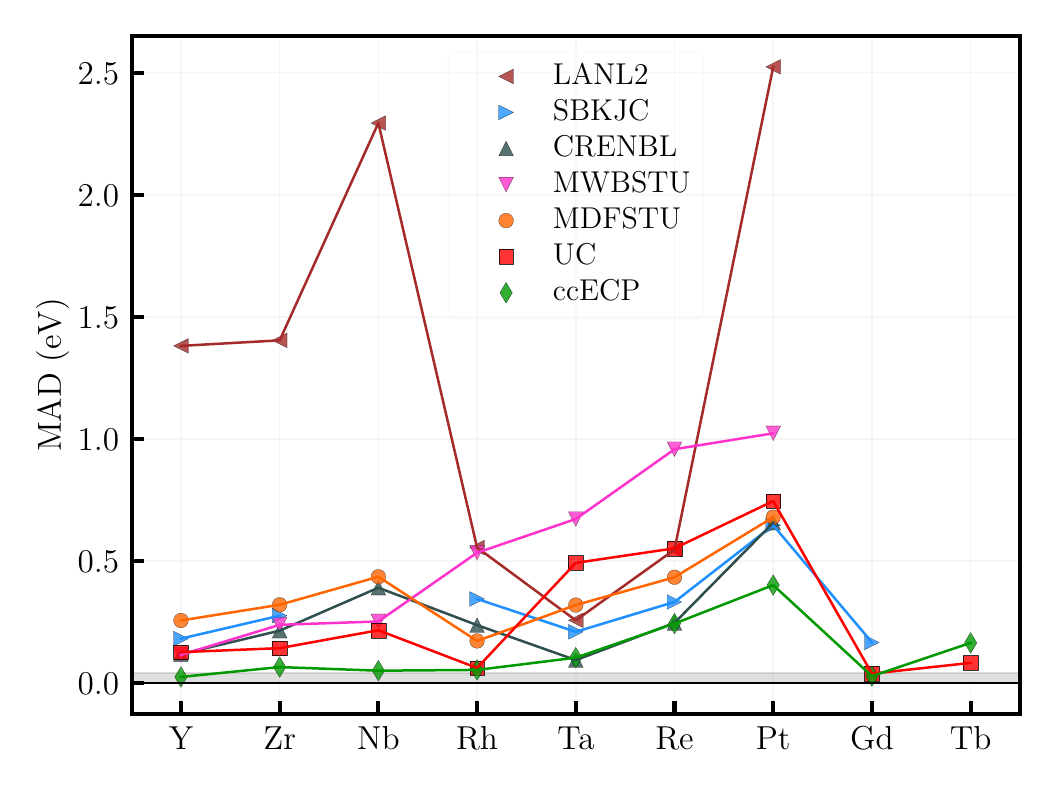}
\caption{
Scalar relativistic AE gap MADs for various core approximations using RCCSD(T) method. 
}
\label{fig:MAD_in_elements}
\end{figure}

\begin{figure}[!htbp]
\centering
\includegraphics[width=1.00\columnwidth]{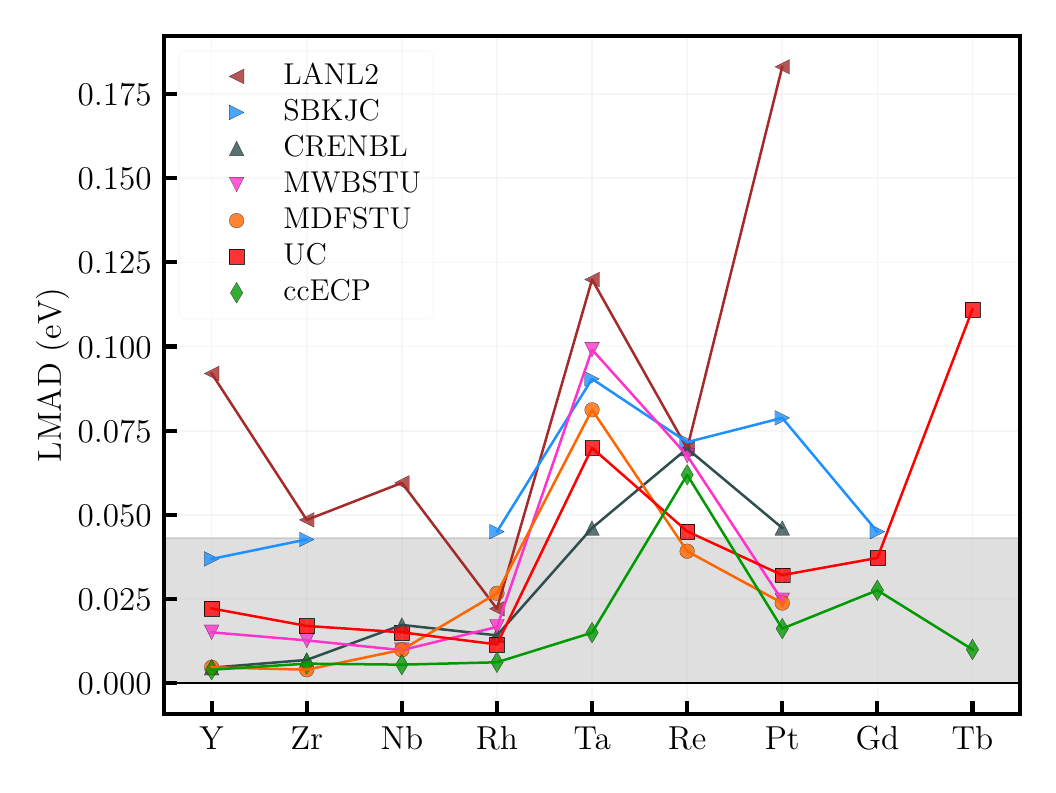}
\caption{
Scalar relativistic AE gap LMADs for various core approximations using RCCSD(T) method. 
}
\label{fig:LMAD_in_elements}
\end{figure}

\begin{figure}[!htbp]
\centering
\includegraphics[width=1.00\columnwidth]{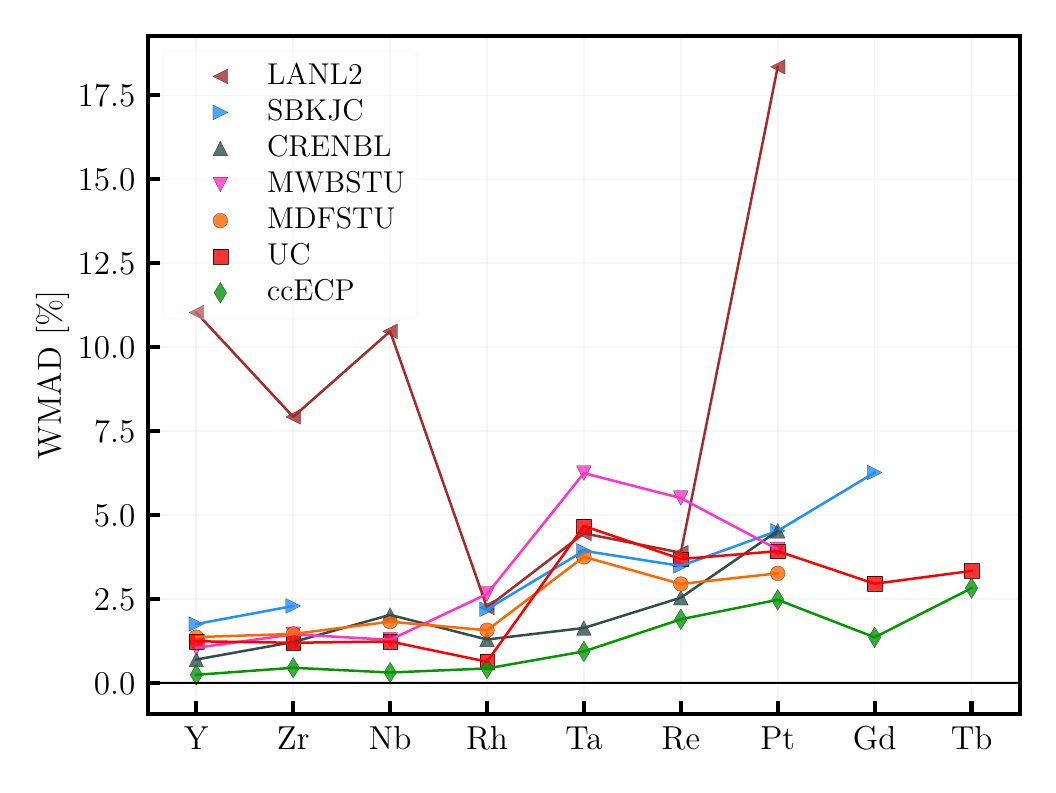}
\caption{
Scalar relativistic AE gap WMADs for various core approximations using RCCSD(T) method.
}
\label{fig:WMAD_in_elements}
\end{figure}

Besides our ccECPs, we calculate the same metrics for various other legacy ECPs for comparison.
These ECPs include MWBSTU\cite{ANDRAE1990}, MDFSTU\cite{Peterson2007, Figgen2009}, CRENBL\cite{LAJOHN1987, R.B.1990}, SBKJC\cite{Stevens1992, Cundari1993}, and LANL2\cite{Wadt1985}.
Additionally, we present the results for uncorrelated core (UC) with correlated treatment restricted to valence subspace only (frozen-core). 
According to the above three defined metrics, MAD, LMAD, and WMAD, the summary of spectral comparison for each element is shown in Figures \ref{fig:MAD_in_elements}, \ref{fig:LMAD_in_elements}, and \ref{fig:WMAD_in_elements}.

For each element, we provide the transferability test of molecular binding energies for all core approximations compared to AE calculation with the fully correlated cores. In order to evaluate the quality of the ECPs, we utilize the discrepancy in dimer binding energies between the ECPs and AE calculations as given by
\begin{equation}
\label{mol_discrep}
    \Delta(r) = D^{ECP}(r) - D^{AE}(r)
\end{equation}
where the $D(r)$ is the binding energy of dimer molecules at the atomic separation length $r$, which ranges from compressed bond length to stretched bond length.
For each core approximation, the calculated binding energies are fitted to the Morse potential:
\begin{equation}
\label{morse_pot}
    V(r)=D_{e}\left(e^{-2 a\left(r-r_{e}\right)}-2 e^{-a\left(r-r_{e}\right)}\right),
\end{equation}
where $D_{e}$ denotes the dissociation energy, $r_e$ labels the equilibrium bond length, and $a$ is a fitting parameter depends on the vibration frequency:
\begin{equation}
    \omega_{e}=\sqrt{\frac{2 a^{2} D_{e}}{\mu}},
\end{equation}
where $\mu$ is the reduced mass of the molecule.

Table \ref{tab:selected_4d_params}, Table \ref{tab:selected_5d_params}, and Table \ref{tab:selected_4f_params} display the optimized parameters for the ccECPs pertaining to the selected elements of $4d$, $5d$, and $4f$ series, respectively.

\begin{table*}
\small
\centering
\caption{Parameters for the selected $4d$ atomic ccECPs. The highest $\ell$ value corresponds to the local channel $L$.
The highest non-local angular momentum channel $\ell_{max}$ is related as $\ell_{max}=L-1$.
}
\label{tab:selected_4d_params}
\begin{tabular}{ccrrrrrccrrrrrrr}
\hline\hline
\multicolumn{1}{c}{Atom} & \multicolumn{1}{c}{$Z_{\rm eff}$} & \multicolumn{1}{c}{Hamiltonian} & \multicolumn{1}{c}{$\ell$} & \multicolumn{1}{c}{$n_{\ell k}$} & \multicolumn{1}{c}{$\alpha_{\ell k}$} & \multicolumn{1}{c}{$\beta_{\ell k}$} & & \multicolumn{1}{c}{Atom} & \multicolumn{1}{c}{$Z_{\rm eff}$} & \multicolumn{1}{c}{Hamiltonian} & \multicolumn{1}{c}{$\ell$} & \multicolumn{1}{c}{$n_{\ell k}$} & \multicolumn{1}{c}{$\alpha_{\ell k}$} & \multicolumn{1}{c}{$\beta_{\ell k}$} \\
\hline
Y  & 11 & AREP &  0 & 2 &    6.868325 &  154.159199  && Zr & 12 & AREP &  0 & 2 &    8.765091 &  150.262314  \\
   &    &      &  0 & 2 &    3.830900 &   18.389590  &&    &    &      &  0 & 2 &    3.928322 &   30.213713  \\
   &    &      &  1 & 2 &    6.193744 &  106.095839  &&    &    &      &  1 & 2 &    7.417907 &   99.301910  \\
   &    &      &  1 & 2 &    2.739296 &   10.237893  &&    &    &      &  1 & 2 &    3.768733 &   25.356026  \\
   &    &      &  2 & 2 &    5.301404 &   48.146429  &&    &    &      &  2 & 2 &    2.250556 &    5.625707  \\
   &    &      &  2 & 2 &    2.236852 &    7.592843  &&    &    &      &  2 & 2 &    5.355356 &   46.916833  \\
   &    &      &  3 & 1 &   11.322164 &   11.000000  &&    &    &      &  3 & 1 &   23.999012 &   12.000000  \\
   &    &      &  3 & 3 &   15.695824 &  124.543799  &&    &    &      &  3 & 3 &   25.971764 &  287.988139  \\
   &    &      &  3 & 2 &    5.067912 &  -20.049244  &&    &    &      &  3 & 2 &    4.926614 &   -5.987178  \\
   &    &      &  3 & 2 &    4.444526 &   -5.929487  &&    &    &      &  3 & 2 &    4.801388 &   -7.922580  \\
   &    &      &    &   &             &              &&    &    &      &    &   &             &              \\
   &    &SOREP &  1 & 2 &    6.789339 &  -58.503206  &&    &    &SOREP &  1 & 2 &    7.660616 &  -58.519285  \\
   &    &      &  1 & 2 &    6.773070 &   58.508207  &&    &    &      &  1 & 2 &    7.440442 &   58.492385  \\
   &    &      &  1 & 2 &    3.078466 &   -7.554229  &&    &    &      &  1 & 2 &    3.308537 &   -7.550185  \\
   &    &      &  1 & 2 &    2.896354 &    7.684825  &&    &    &      &  1 & 2 &    3.147063 &    7.738905  \\
   &    &      &  2 & 2 &    5.419086 &  -11.851193  &&    &    &      &  2 & 2 &    5.481717 &  -11.708397  \\
   &    &      &  2 & 2 &    5.333864 &   11.852624  &&    &    &      &  2 & 2 &    6.090472 &   11.867706  \\
   &    &      &  2 & 2 &    1.970651 &   -2.065575  &&    &    &      &  2 & 2 &    2.361902 &   -2.486771  \\
   &    &      &  2 & 2 &    1.954057 &    2.054659  &&    &    &      &  2 & 2 &    1.764896 &    1.566729  \\
   &    &      &    &   &             &              &&    &    &      &    &   &             &              \\
Nb & 13 & AREP &  0 & 2 &    8.451450 &  188.147277  && Rh & 17 & AREP &  0 & 2 &   11.968516 &  245.387919  \\
   &    &      &  0 & 2 &    4.655854 &   25.673881  &&    &    &      &  0 & 2 &    5.125760 &   27.285840  \\
   &    &      &  1 & 2 &    8.192024 &  134.298070  &&    &    &      &  1 & 2 &    9.739600 &  180.589423  \\
   &    &      &  1 & 2 &    3.466263 &   17.381556  &&    &    &      &  1 & 2 &    6.564776 &   28.411628  \\
   &    &      &  2 & 2 &    6.825072 &   60.151219  &&    &    &      &  2 & 2 &   11.425921 &   83.487044  \\
   &    &      &  2 & 2 &    2.858157 &   10.393369  &&    &    &      &  2 & 2 &    4.502521 &   21.380372  \\
   &    &      &  3 & 1 &   19.860892 &   13.000000  &&    &    &      &  3 & 1 &   22.806206 &   17.000000  \\
   &    &      &  3 & 3 &   19.525927 &  258.191591  &&    &    &      &  3 & 3 &   32.710006 &  387.705494  \\
   &    &      &  3 & 2 &    7.388201 &  -23.263128  &&    &    &      &  3 & 2 &   14.048587 &  -10.571967  \\
   &    &      &  3 & 2 &    4.558641 &   -4.887375  &&    &    &      &  3 & 2 &   11.838335 &  -14.594461  \\
   &    &      &    &   &             &              &&    &    &      &    &   &             &              \\
   &    &SOREP &  1 & 2 &    8.359425 &  -74.498506  &&    &    & SOREP&  1 & 2 &   12.583835 & -105.691011  \\
   &    &      &  1 & 2 &    8.167350 &   74.507443  &&    &    &      &  1 & 2 &    9.850973 &  105.807831  \\
   &    &      &  1 & 2 &    3.685605 &  -10.879317  &&    &    &      &  1 & 4 &    4.165404 &  -17.209011  \\
   &    &      &  1 & 2 &    3.551957 &   10.916537  &&    &    &      &  1 & 4 &    6.179692 &   17.035664  \\
   &    &      &  2 & 2 &    6.688623 &  -15.214533  &&    &    &      &  2 & 2 &   12.428318 &  -24.301306  \\
   &    &      &  2 & 2 &    6.537684 &   15.220185  &&    &    &      &  2 & 2 &   14.006097 &   24.821099  \\
   &    &      &  2 & 2 &    2.551296 &   -3.000998  &&    &    &      &  2 & 2 &    4.782894 &   -2.724934  \\
   &    &      &  2 & 2 &    2.557909 &    3.036924  &&    &    &      &  2 & 2 &    5.952595 &    6.111212  \\
   &    &      &    &   &             &              &&    &    &      &    &   &             &              \\
\hline\hline
\end{tabular}
\end{table*}

\begin{table*}
\small
\centering
\caption{
Parameters for the selected $5d$ atomic ccECPs. The highest $\ell$ value corresponds to the local channel $L$.
The highest non-local angular momentum channel $\ell_{max}$ is related as $\ell_{max}=L-1$.
}
\label{tab:selected_5d_params}
\begin{tabular}{ccrrrrrccrrrrrrr}
\hline\hline
\multicolumn{1}{c}{Atom} & \multicolumn{1}{c}{$Z_{\rm eff}$} & \multicolumn{1}{c}{Hamiltonian} & \multicolumn{1}{c}{$\ell$} & \multicolumn{1}{c}{$n_{\ell k}$} & \multicolumn{1}{c}{$\alpha_{\ell k}$} & \multicolumn{1}{c}{$\beta_{\ell k}$} & & \multicolumn{1}{c}{Atom} & \multicolumn{1}{c}{$Z_{\rm eff}$} & \multicolumn{1}{c}{Hamiltonian} & \multicolumn{1}{c}{$\ell$} & \multicolumn{1}{c}{$n_{\ell k}$} & \multicolumn{1}{c}{$\alpha_{\ell k}$} & \multicolumn{1}{c}{$\beta_{\ell k}$} \\
\hline
Ta & 13 & AREP &  0 & 2 &   16.084877 &   354.560780 && Re & 15 & AREP &  0 & 2 &   11.527415 &  471.040560  \\
   &    &      &  0 & 4 &    4.856689 &    -3.189062 &&    &    &      &  0 & 2 &    3.521939 &   17.808887  \\
   &    &      &  0 & 2 &    6.308029 &    15.434060 &&    &    &      &  1 & 2 &    9.670534 &  265.269318  \\
   &    &      &  0 & 2 &    4.049118 &    17.921890 &&    &    &      &  1 & 2 &    4.693214 &   48.515418  \\
   &    &      &  1 & 2 &    9.219236 &   291.708719 &&    &    &      &  2 & 2 &    6.202607 &  107.920979  \\
   &    &      &  1 & 4 &    8.304225 &   -10.543181 &&    &    &      &  2 & 2 &    4.054883 &   31.437969  \\
   &    &      &  1 & 2 &    3.160451 &    -0.606327 &&    &    &      &  3 & 2 &    2.502922 &   16.905644  \\
   &    &      &  1 & 2 &    2.276207 &     3.707560 &&    &    &      &  3 & 2 &    4.013976 &   17.858142  \\
   &    &      &  2 & 2 &    5.663931 &   119.434072 &&    &    &      &  4 & 1 &   13.992911 &   15.000000  \\
   &    &      &  2 & 4 &    6.804907 &    -1.428753 &&    &    &      &  4 & 3 &   13.164906 &  209.893668  \\
   &    &      &  2 & 2 &    5.867466 &     6.084632 &&    &    &      &  4 & 2 &    3.873316 &   -8.076852  \\
   &    &      &  2 & 2 &    5.458504 &    14.589293 &&    &    &      &  4 & 2 &    3.654226 &   -9.999694  \\
   &    &      &  3 & 2 &    1.885697 &    13.741560 &&    &    &      &    &   &             &              \\
   &    &      &  3 & 2 &    3.231234 &    12.967039 &&    &    & SOREP&  1 & 2 &    9.196200 & -176.980000  \\
   &    &      &  4 & 1 &    6.585798 &    13.000000 &&    &    &      &  1 & 2 &    9.692400 &  176.770000  \\
   &    &      &  4 & 3 &    9.131928 &    85.615374 &&    &    &      &  1 & 2 &    7.722000 &  -20.947000  \\
   &    &      &  4 & 2 &    7.991861 &    -0.262937 &&    &    &      &  1 & 2 &    3.956400 &   20.428000  \\
   &    &      &  4 & 2 &    1.842639 &    -2.273914 &&    &    &      &  2 & 2 &    6.157000 &  -43.309000  \\
   &    &      &    &   &             &              &&    &    &      &  2 & 2 &    6.201600 &   43.086000  \\
   &    & SOREP&  1 & 2 &    9.567757 &  -193.788022 &&    &    &      &  2 & 2 &    4.608100 &   -5.345500  \\
   &    &      &  1 & 2 &    7.672453 &   195.894081 &&    &    &      &  2 & 2 &    3.941100 &    5.715900  \\
   &    &      &  1 & 4 &    9.284590 &    -9.653330 &&    &    &      &  3 & 2 &    2.562700 &   -4.830900  \\
   &    &      &  1 & 4 &    8.099739 &     6.253216 &&    &    &      &  3 & 2 &    2.521500 &    4.830200  \\
   &    &      &  1 & 2 &    3.637006 &    -0.144229 &&    &    &      &    &   &             &              \\
   &    &      &  1 & 2 &    1.690943 &    -1.601960 && Pt & 18 & AREP &  0 & 2 &   14.604490 &  428.400991  \\
   &    &      &  2 & 2 &    5.498582 &   -45.969268 &&    &    &      &  0 & 2 &    7.218055 &   67.391235  \\
   &    &      &  2 & 2 &    5.595047 &    46.401258 &&    &    &      &  0 & 2 &    5.429571 &   21.842450  \\
   &    &      &  2 & 4 &    5.868634 &    -0.474343 &&    &    &      &  1 & 2 &   11.178990 &  255.947773  \\
   &    &      &  2 & 4 &    5.654115 &     0.293785 &&    &    &      &  1 & 2 &    5.365413 &   52.709385  \\
   &    &      &  2 & 2 &    1.414594 &     0.484404 &&    &    &      &  1 & 2 &    8.011118 &   13.379362  \\
   &    &      &  2 & 2 &    1.565152 &    -0.270081 &&    &    &      &  2 & 2 &    7.615605 &  121.977058  \\
   &    &      &  3 & 2 &    2.132660 &    -3.831108 &&    &    &      &  2 & 2 &    3.904912 &    5.839996  \\
   &    &      &  3 & 2 &    2.032923 &     3.924549 &&    &    &      &  2 & 2 &    5.446875 &   44.746892  \\
   &    &      &    &   &             &              &&    &    &      &  3 & 2 &    3.386257 &   13.308218  \\
   &    &      &    &   &             &              &&    &    &      &  3 & 2 &    3.326489 &   19.517395  \\
   &    &      &    &   &             &              &&    &    &      &  3 & 2 &    5.430616 &   25.350338  \\
   &    &      &    &   &             &              &&    &    &      &  4 & 1 &   13.600000 &   18.000000  \\
   &    &      &    &   &             &              &&    &    &      &  4 & 3 &   13.600000 &  244.800000  \\
   &    &      &    &   &             &              &&    &    &      &  4 & 2 &   13.600000 & -161.707117  \\
   &    &      &    &   &             &              &&    &    &      &  4 & 2 &    5.430186 &  -19.530044  \\
   &    &      &    &   &             &              &&    &    &      &    &   &             &              \\
   &    &      &    &   &             &              &&    &    &SOREP &  1 & 2 &   12.905978 & -176.006816  \\
   &    &      &    &   &             &              &&    &    &      &  1 & 2 &    9.569534 &  176.041345  \\
   &    &      &    &   &             &              &&    &    &      &  1 & 2 &    5.241937 &  -27.260853  \\
   &    &      &    &   &             &              &&    &    &      &  1 & 2 &    7.370807 &   27.417143  \\
   &    &      &    &   &             &              &&    &    &      &  2 & 2 &    8.156923 &  -42.598141  \\
   &    &      &    &   &             &              &&    &    &      &  2 & 2 &    7.160691 &   44.200769  \\
   &    &      &    &   &             &              &&    &    &      &  2 & 2 &    3.611191 &   -6.876859  \\
   &    &      &    &   &             &              &&    &    &      &  2 & 2 &    4.034286 &    7.580091  \\
   &    &      &    &   &             &              &&    &    &      &  3 & 2 &    3.379869 &   -7.140146  \\
   &    &      &    &   &             &              &&    &    &      &  3 & 2 &    3.326255 &    7.139063  \\
   &    &      &    &   &             &              &&    &    &      &    &   &             &              \\
\hline\hline
\end{tabular}
\end{table*}

\begin{table*}
\small
\centering
\caption{Parameters for the selected $4f$ atomic ccECPs. The highest $\ell$ value corresponds to the local channel $L$.
The highest non-local angular momentum channel $\ell_{max}$ is related as $\ell_{max}=L-1$.
}
\label{tab:selected_4f_params}
\begin{tabular}{ccrrrrrccrrrrrrr}
\hline\hline
\multicolumn{1}{c}{Atom} & \multicolumn{1}{c}{$Z_{\rm eff}$} & \multicolumn{1}{c}{Hamiltonian} & \multicolumn{1}{c}{$\ell$} & \multicolumn{1}{c}{$n_{\ell k}$} & \multicolumn{1}{c}{$\alpha_{\ell k}$} & \multicolumn{1}{c}{$\beta_{\ell k}$} & & \multicolumn{1}{c}{Atom} & \multicolumn{1}{c}{$Z_{\rm eff}$} & \multicolumn{1}{c}{Hamiltonian} & \multicolumn{1}{c}{$\ell$} & \multicolumn{1}{c}{$n_{\ell k}$} & \multicolumn{1}{c}{$\alpha_{\ell k}$} & \multicolumn{1}{c}{$\beta_{\ell k}$} \\
\hline
Gd & 18 & AREP &  0 & 2 &   11.958743 &  158.751573  && Tb & 19 & AREP &  0 & 2 &    8.996757 &  161.119164   \\
   &    &      &  0 & 2 &    4.575020 &   75.338628  &&    &    &      &  0 & 2 &    5.104371 &   89.334835   \\
   &    &      &  1 & 2 &    9.327027 &   85.299703  &&    &    &      &  1 & 2 &    7.640657 &   88.693021   \\
   &    &      &  1 & 2 &    3.727309 &   55.148896  &&    &    &      &  1 & 2 &    3.878305 &   57.109488   \\
   &    &      &  2 & 2 &    9.142398 &   55.315114  &&    &    &      &  2 & 2 &    7.456305 &   55.901690   \\
   &    &      &  2 & 2 &    2.410263 &   27.587515  &&    &    &      &  2 & 2 &    2.530445 &   29.364333   \\
   &    &      &  3 & 2 &    3.496091 &  -24.952915  &&    &    &      &  3 & 2 &    4.939345 &  -23.952530   \\
   &    &      &  3 & 2 &    6.162299 &   -2.920504  &&    &    &      &  3 & 2 &    3.912490 &   -7.090915   \\
   &    &      &  4 & 1 &    4.482335 &   18.000000  &&    &    &      &  4 & 1 &    4.526254 &   19.000000   \\
   &    &      &  4 & 3 &    4.564890 &   82.168015  &&    &    &      &  4 & 3 &    3.879308 &   85.998831   \\
   &    &      &  4 & 2 &    5.667120 &  -92.949026  &&    &    &      &  4 & 2 &    4.406058 &  -85.722649   \\
   &    &      &  4 & 2 &    3.161461 &   -4.588504  &&    &    &      &  4 & 2 &    1.899664 &   -3.133952   \\
   &    &      &    &   &             &              &&    &    &      &    &   &             &               \\
   &    & SOREP&  1 & 4 &    2.002999 &    2.782860  &&    &    & SOREP&  1 & 4 &    2.080465 &    2.963688   \\
   &    &      &  2 & 4 &    4.114860 &   -0.322458  &&    &    &      &  2 & 4 &    2.660751 &    0.922456   \\
   &    &      &  3 & 4 &    4.965905 &    0.231259  &&    &    &      &  3 & 4 &    4.708055 &    0.178089   \\
   &    &      &    &   &             &              &&    &    &      &    &   &             &               \\
\hline\hline
\end{tabular}
\end{table*}

\subsection{Selected $4d$ elements}
In this section, we discuss the results for Y, Zr, Nb, and Rh. The elements in this section use the same 28 electron core [Ar]$3d^{10}$ as in our previous ccECP work \cite{Wang2022} with $s$, $p$, and $d$ non-local channels. 
The SBKJC ECP was excluded for Nb in both atomic and molecular comparisons due to difficulties with CCSD(T) convergence.
At the AREP level, the constructed ccECPs show significant improvement in terms of the accuracy of atomic spectra as shown in Figure \ref{fig:MAD_in_elements}, \ref{fig:LMAD_in_elements}, and \ref{fig:WMAD_in_elements}.

\begin{figure*}[!htbp]
\centering
\begin{subfigure}{0.5\textwidth}
\includegraphics[width=\textwidth]{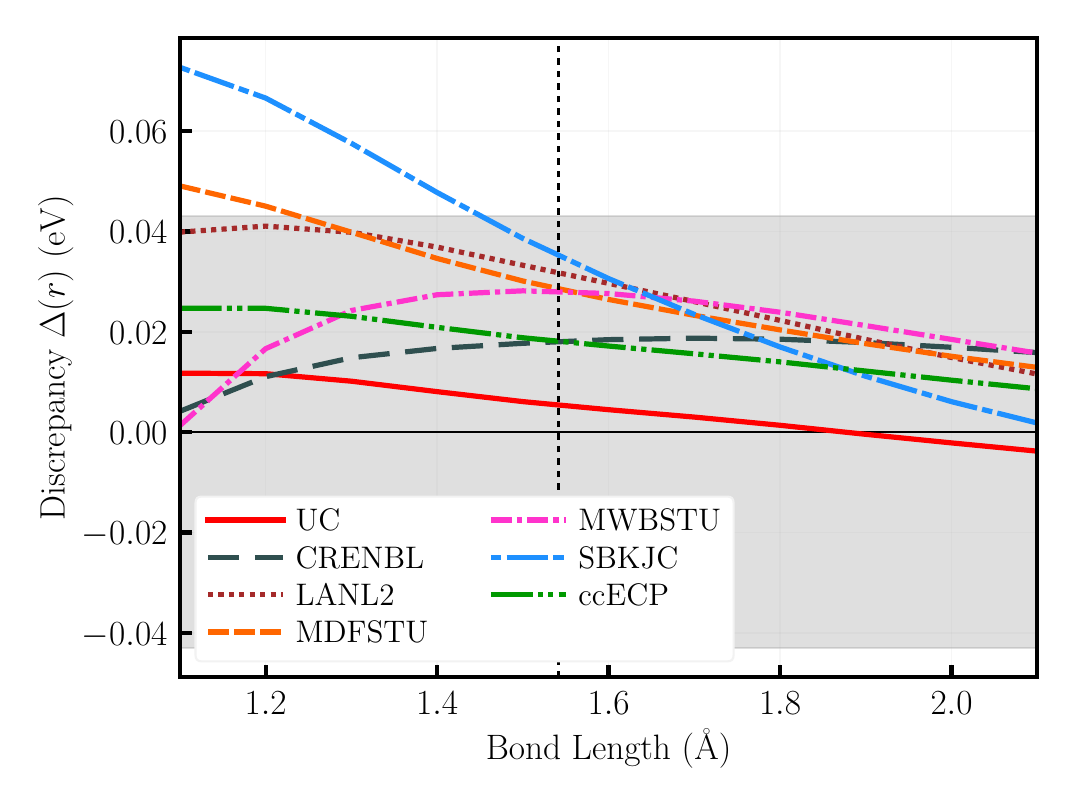}
\caption{RhH binding curve discrepancies}
\label{fig:RhH}
\end{subfigure}%
\begin{subfigure}{0.5\textwidth}
\includegraphics[width=\textwidth]{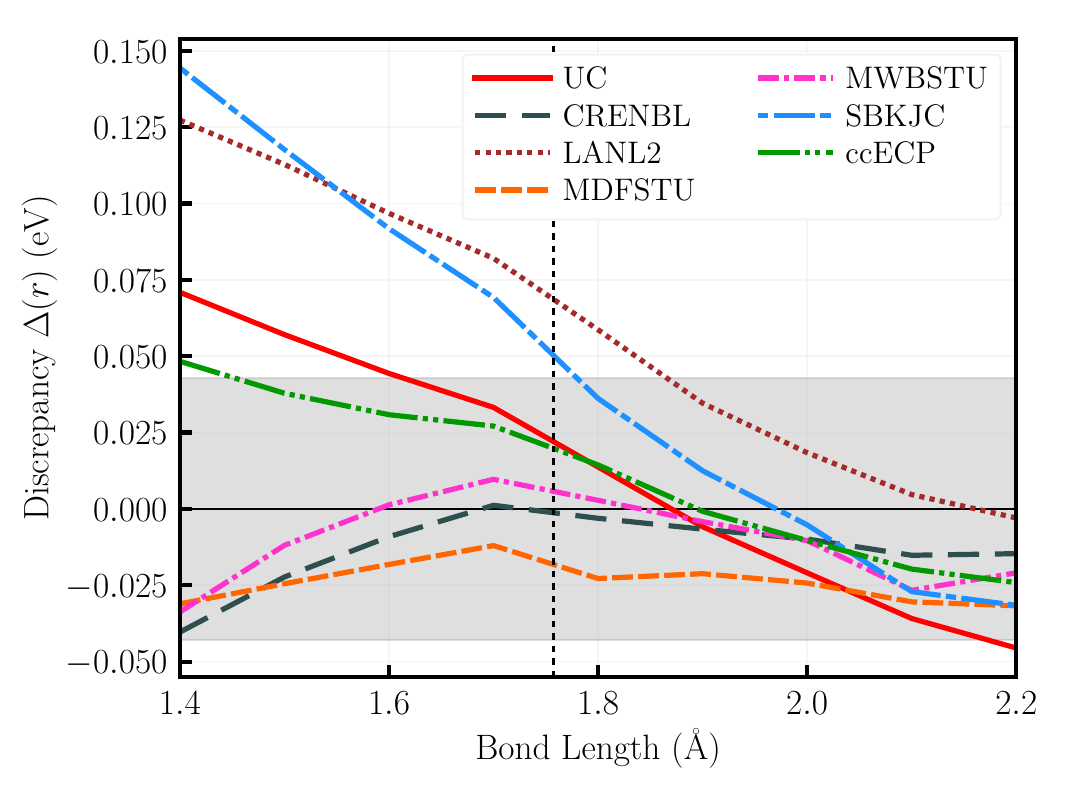}
\caption{RhO binding curve discrepancies}
\label{fig:RhO}
\end{subfigure}
\caption{Binding energy discrepancies for (a) RhH and (b) RhO molecules. The grey-shaded region indicates the chemical accuracy, and the vertical dashed line gives the equilibrium bond length from the AE result.}
\label{fig:Rh_mols}
\end{figure*}

The binding energy discrepancy is plotted in a wide range of bond lengths, from near the dissociation limit at short distances to stretched bond lengths.
We observe modest improvement in the binding curves of monohydrides and monoxides for these selected $4d$ elements. 
The results are shown in the following Figures \ref{fig:Y_mols}, \ref{fig:Zr_mols}, \ref{fig:Nb_mols}, and \ref{fig:Rh_mols}.

\begin{figure*}[!htbp]
\centering
\begin{subfigure}{0.5\textwidth}
\includegraphics[width=\textwidth]{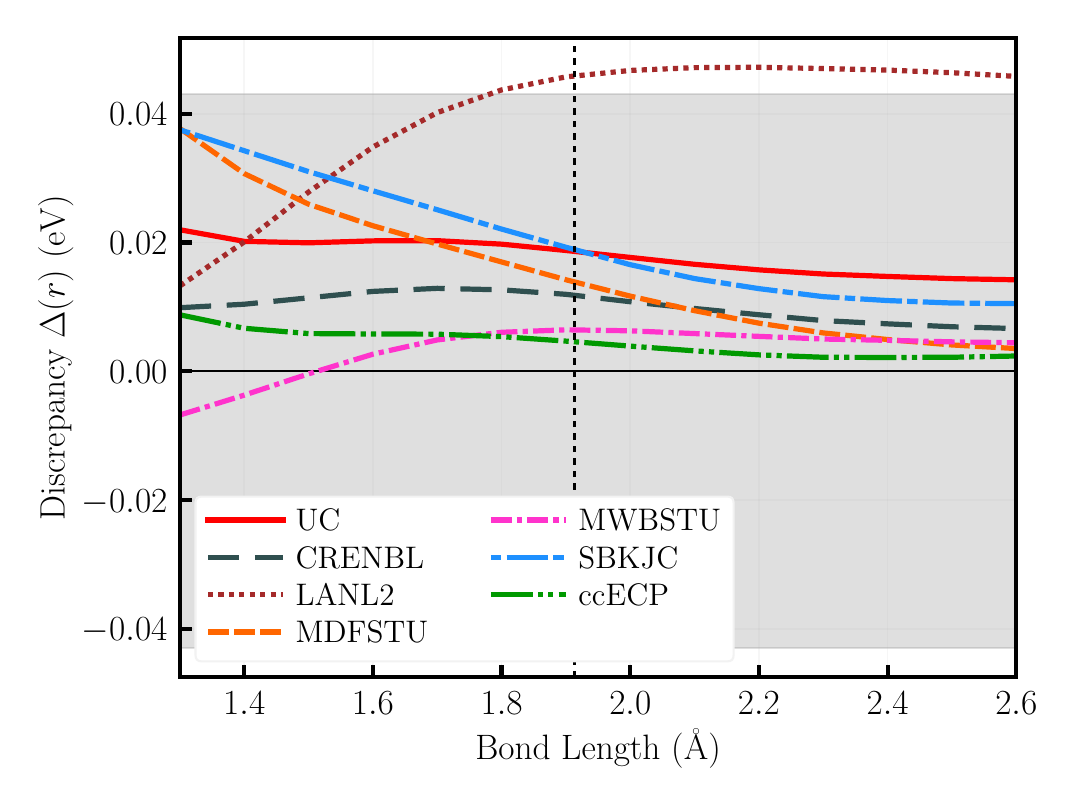}
\caption{YH binding curve discrepancies}
\label{fig:YH}
\end{subfigure}%
\begin{subfigure}{0.5\textwidth}
\includegraphics[width=\textwidth]{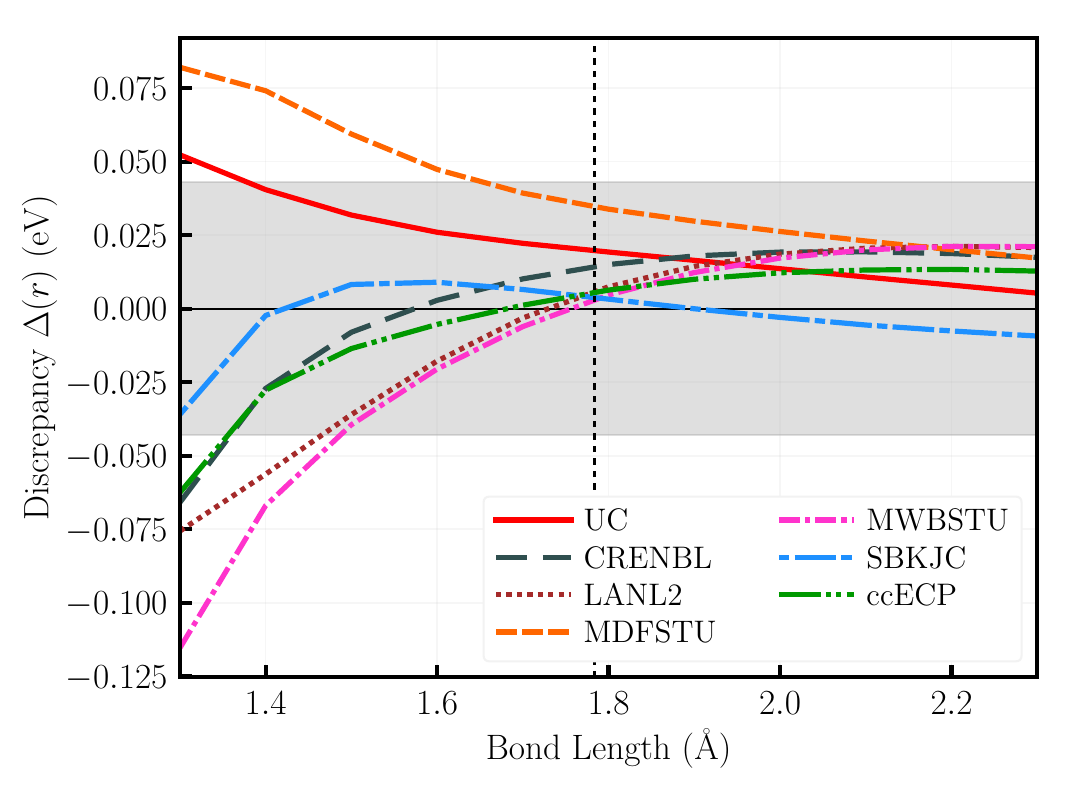}
\caption{YO binding curve discrepancies}
\label{fig:YO}
\end{subfigure}
\caption{Binding energy discrepancies for (a) YH and (b) YO molecules. The grey-shaded region indicates the chemical accuracy, and the vertical dashed line gives the equilibrium bond length from the AE result.}
\label{fig:Y_mols}
\end{figure*}

\begin{figure*}[!htbp]
\centering
\begin{subfigure}{0.5\textwidth}
\includegraphics[width=\textwidth]{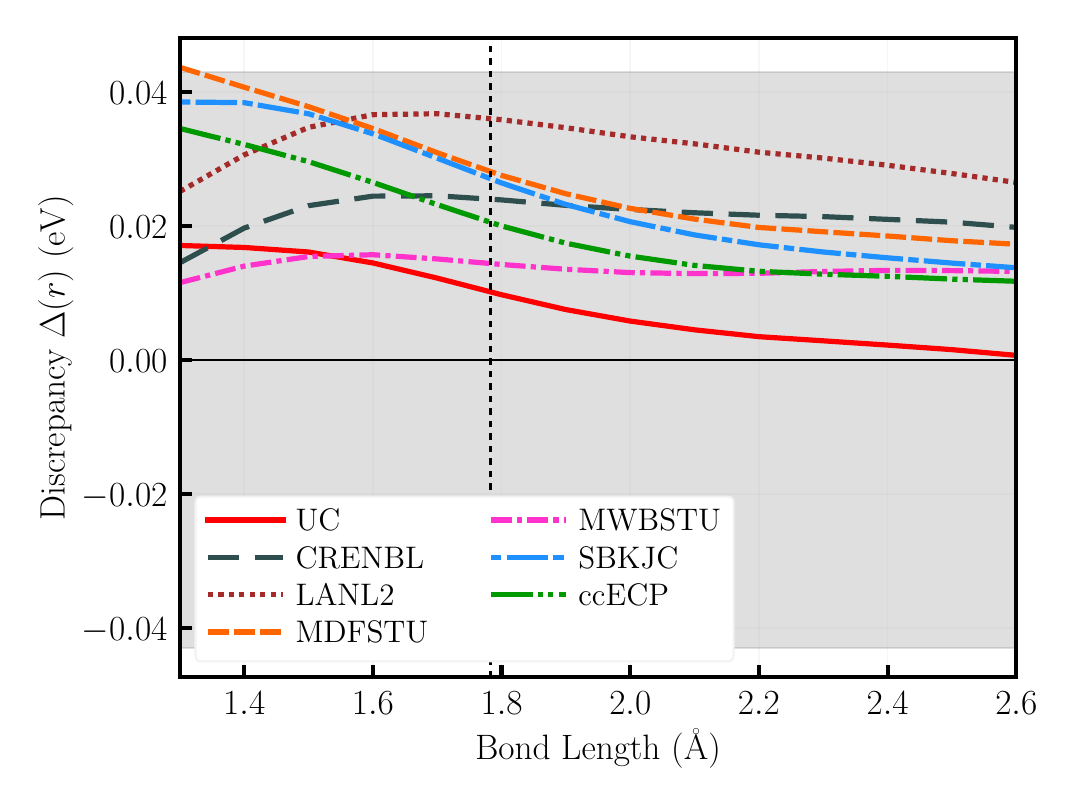}
\caption{ZrH binding curve discrepancies}
\label{fig:ZrH}
\end{subfigure}%
\begin{subfigure}{0.5\textwidth}
\includegraphics[width=\textwidth]{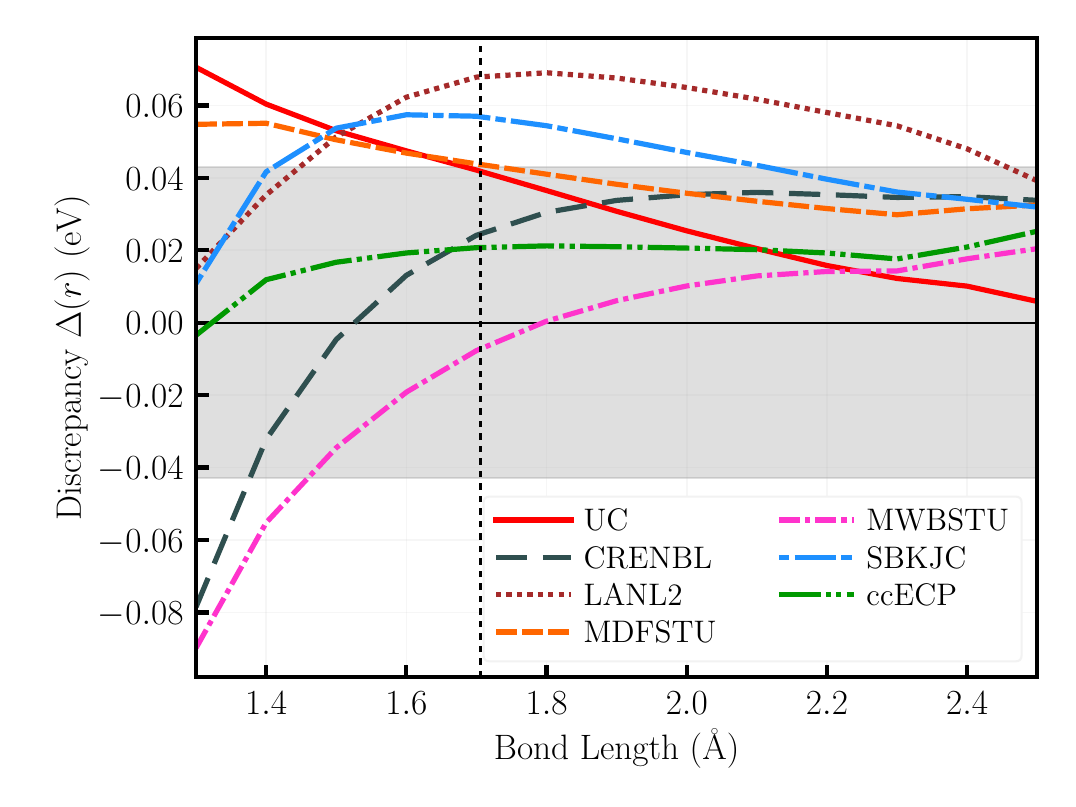}
\caption{ZrO binding curve discrepancies}
\label{fig:ZrO}
\end{subfigure}
\caption{Binding energy discrepancies for (a) ZrH and (b) ZrO molecules. The grey-shaded region indicates the chemical accuracy, and the vertical dashed line gives the equilibrium bond length from the AE result.}
\label{fig:Zr_mols}
\end{figure*}

\begin{figure*}[!htbp]
\centering
\begin{subfigure}{0.5\textwidth}
\includegraphics[width=\textwidth]{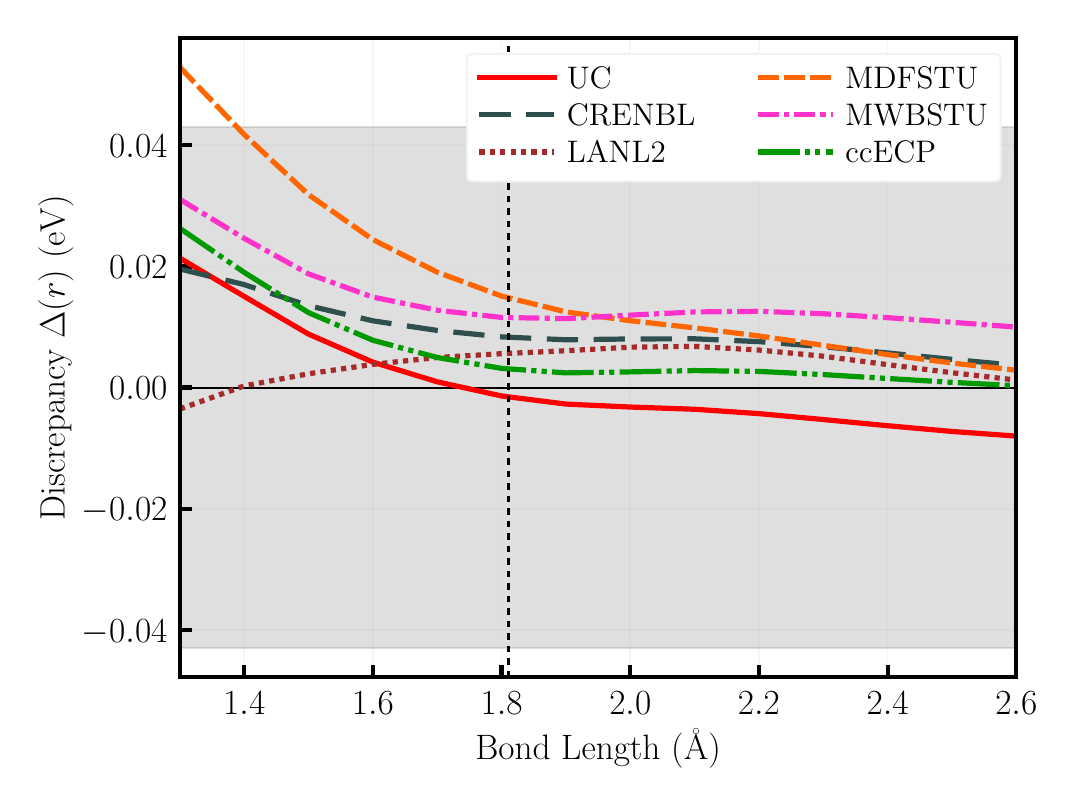}
\caption{NbH binding curve discrepancies}
\label{fig:NbH}
\end{subfigure}%
\begin{subfigure}{0.5\textwidth}
\includegraphics[width=\textwidth]{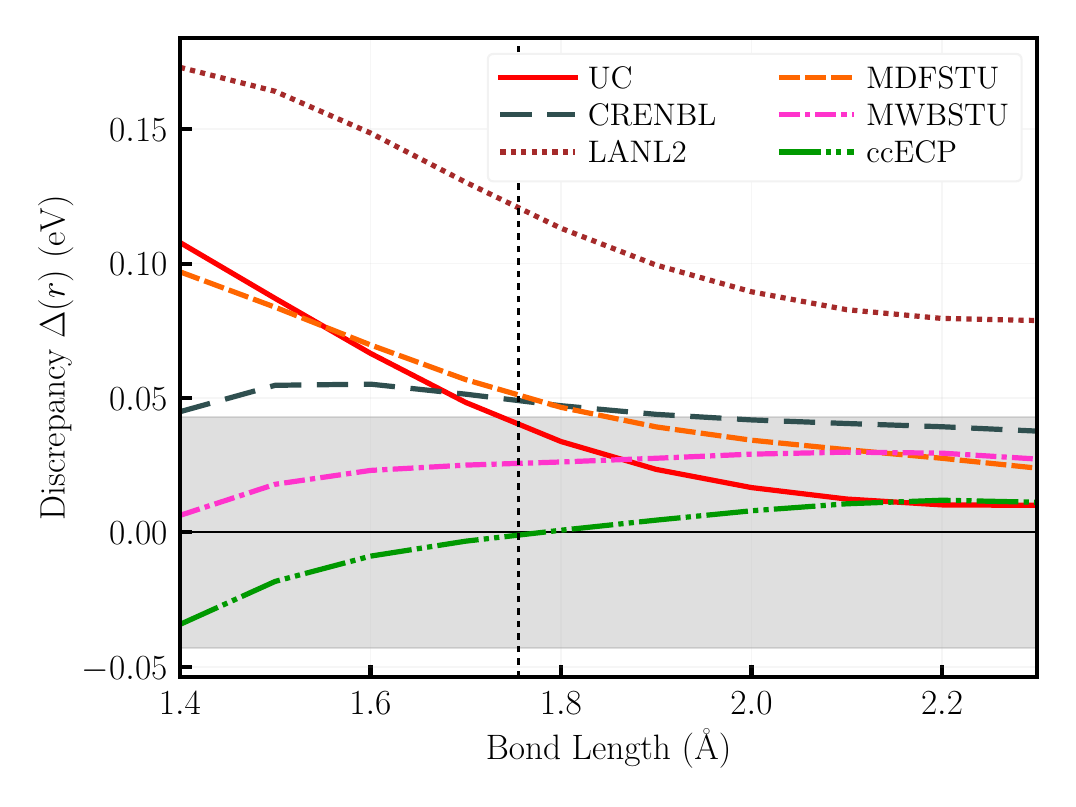}
\caption{NbO binding curve discrepancies}
\label{fig:NbO}
\end{subfigure}
\caption{Binding energy discrepancies for (a) NbH and (b) NbO molecules. The grey-shaded region indicates the chemical accuracy, and the vertical dashed line gives the equilibrium bond length from the AE result.}
\label{fig:Nb_mols}
\end{figure*}

We can observe that for Yttrium, given its simple structure, all the ECPs are presenting binding curves that are almost completely within the chemical accuracy bounds. 
We note that our ccECPs for Y,Zr, Nb, and Rh, show systematic improvements when compared to other ECPs. We see consistently small discrepancies at equilibrium bond lengths as well as throughout the entire bond lengths.
Improvements are more noticeable in the oxide binding scenario, especially in ZrO and NbO, where ccECP outperforms other ECPs significantly to maintain accuracy in all geometries.

Our AREP ccECPs show systematic
improvement when compared to legacy ECPs, both in atomic spectra and molecular tests.
In Table \ref{morse:4d}, we provide the summary of mean absolute deviations for the binding parameters from the fit of the Morse potential. 
ccECP shows overall systematic consistency and a good balance between the imposed optimization criteria.

\begin{table}[!htbp]
\centering
\caption{Mean absolute deviations of molecular binding parameters for various core approximations with respect to AE data for $4d$ selected heavy-element groups, Y, Zr, Nb, and Rh related molecules. All parameters were obtained using Morse potential fit. The parameters shown are dissociation energy $D_e$, equilibrium bond length $r_e$, vibrational frequency $\omega_e$ and binding energy discrepancy at dissociation bond length $D_{diss}$.}
\label{morse:4d}
\begin{tabular}{l|rrrrrrrrrr}
\hline\hline
{} & $D_e$(eV) & $r_e$(\AA) & $\omega_e$(cm$^{-1}$) & $D_{diss}$(eV) \\
\hline
UC      &   0.02(1) &   0.002(2) &                  4(7) &        0.05(8) \\
CRENBL  &   0.02(1) &   0.001(2) &                  2(7) &        0.03(8) \\
LANL2   &   0.05(1) &   0.003(2) &                  6(7) &        0.06(8) \\
MDFSTU  &   0.03(1) &   0.002(2) &                  4(7) &        0.05(8) \\
MWBSTU  &   0.01(1) &   0.001(2) &                  3(7) &        0.03(8) \\
SBKJC   &   0.03(1) &   0.004(2) &                 10(7) &        0.05(7) \\
ccECP   &   0.01(1) &   0.002(2) &                  3(7) &        0.03(8) \\
\hline\hline
\end{tabular}
\end{table}

\subsection{Selected $5d$ elements}

This section includes elements Ta, Re, and Pt. The 60-electron core employed 
(configuration [Kr]$4d^{10}4f^{14}$) is identical to the core used for $5d$ elements from in our previous work \cite{Wang2022}. 


The atomic spectra accuracy are presented with MAD, LMAD, and WMAD plots in Figures \ref{fig:MAD_in_elements}, \ref{fig:LMAD_in_elements}, and \ref{fig:WMAD_in_elements}, respectively. 
We observe consistent improvements in our ccECPs for Ta, Re, and Pt compared to other ECPs and uncorrelated cores. 
For Ta, Re, and Pt, the MADs are clearly better than legacy ECPs.
The LMAD for Tb is a little unique compared to the others shown. 
For Tb we were unable to converge an electron affinity state, so the calculation of the LMAD only includes the first excited state, first ionization potential, and second ionization potential.
 For low-lying states, Ta also shows significant improvement in the sense that all the other ECPs LMADs exceed the chemical accuracy, whereas ccECP achieves a comparable LMAD to that of some lighter transition metals. 
The LMADs for Re and Pt are also on par or better than the previous ECPs. 
Additionally, our ccECPs for Ta, Re, and Pt show substantial improvements in terms of WMAD. 

\begin{figure*}[!htbp]
\centering
\begin{subfigure}{0.5\textwidth}
\includegraphics[width=\textwidth]{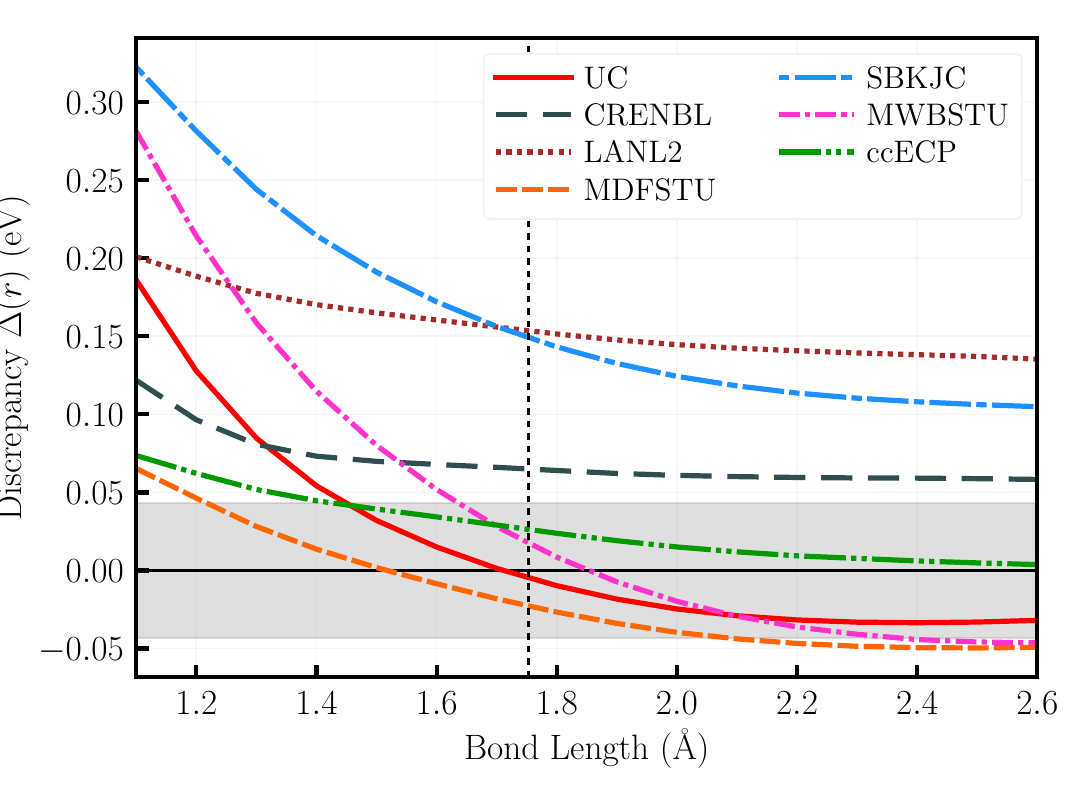}
\caption{TaH binding curve discrepancies}
\label{fig:TaH}
\end{subfigure}%
\begin{subfigure}{0.5\textwidth}
\includegraphics[width=\textwidth]{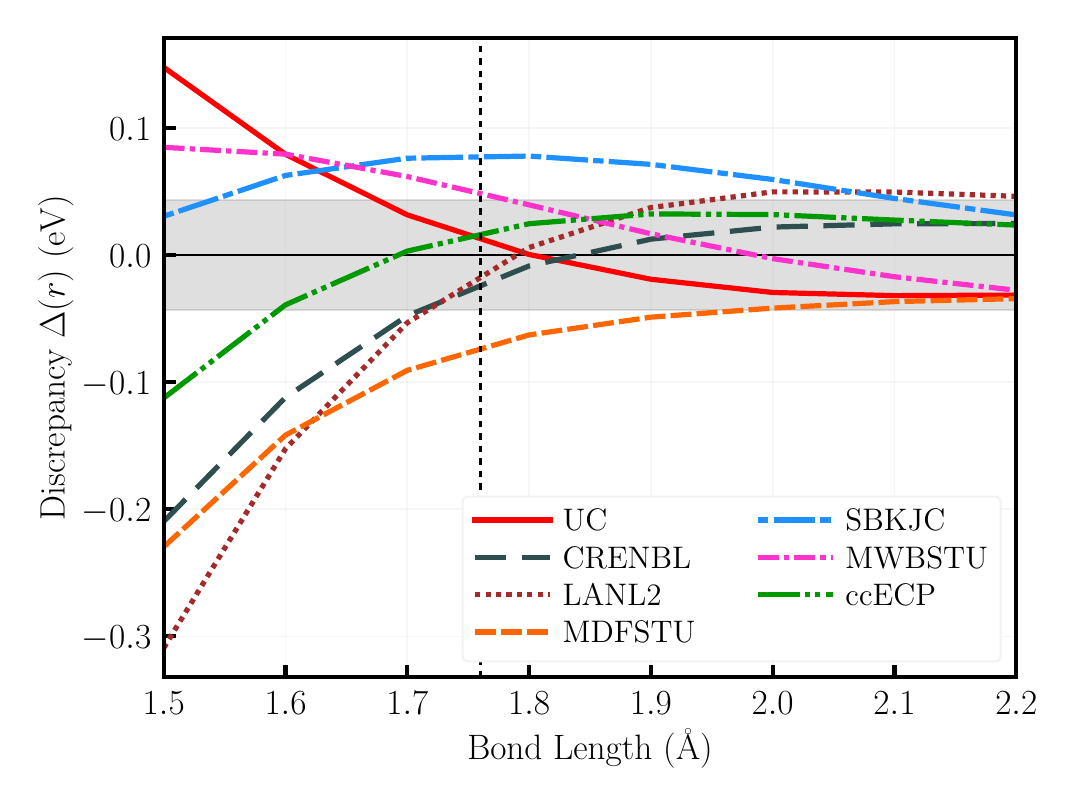}
\caption{TaO binding curve discrepancies}
\label{fig:TaO}
\end{subfigure}
\caption{Binding energy discrepancies for (a) TaH and (b) TaO molecules. The grey-shaded region indicates the chemical accuracy, and the vertical dashed line gives the equilibrium bond length from the AE result.}
\label{fig:Ta_mols}
\end{figure*}

\begin{figure*}[!htbp]
\centering
\begin{subfigure}{0.5\textwidth}
\includegraphics[width=\textwidth]{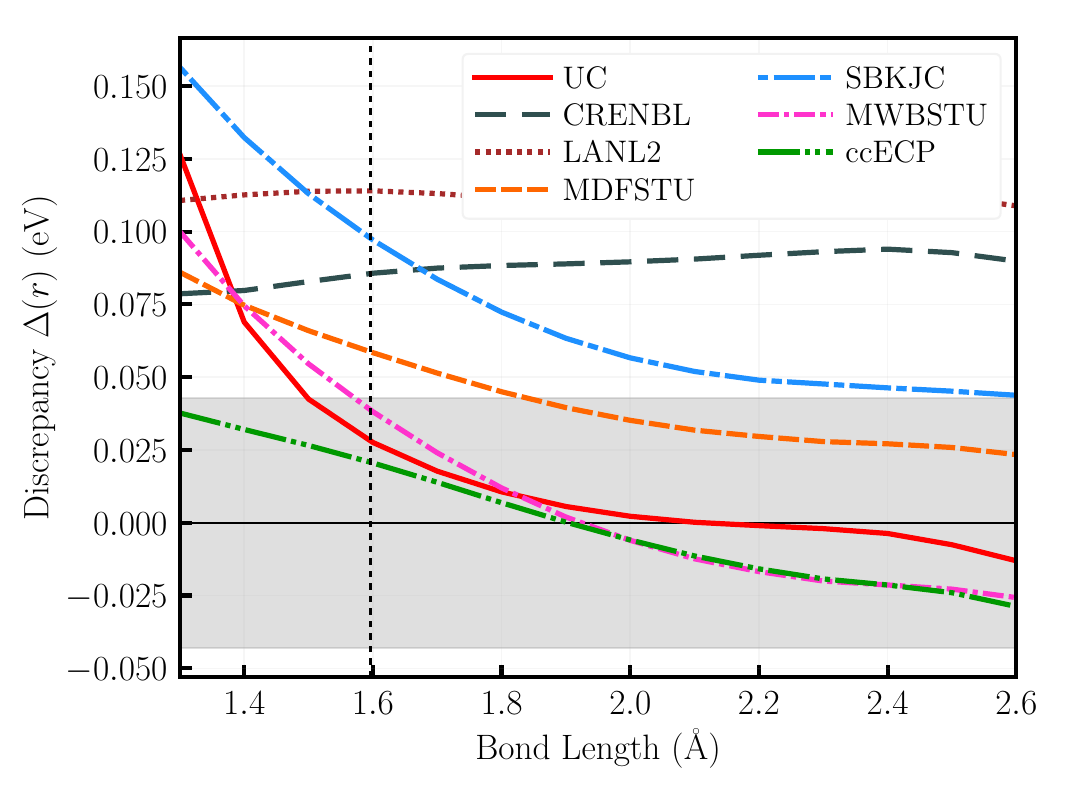}
\caption{ReH binding curve discrepancies}
\label{fig:ReH}
\end{subfigure}%
\begin{subfigure}{0.5\textwidth}
\includegraphics[width=\textwidth]{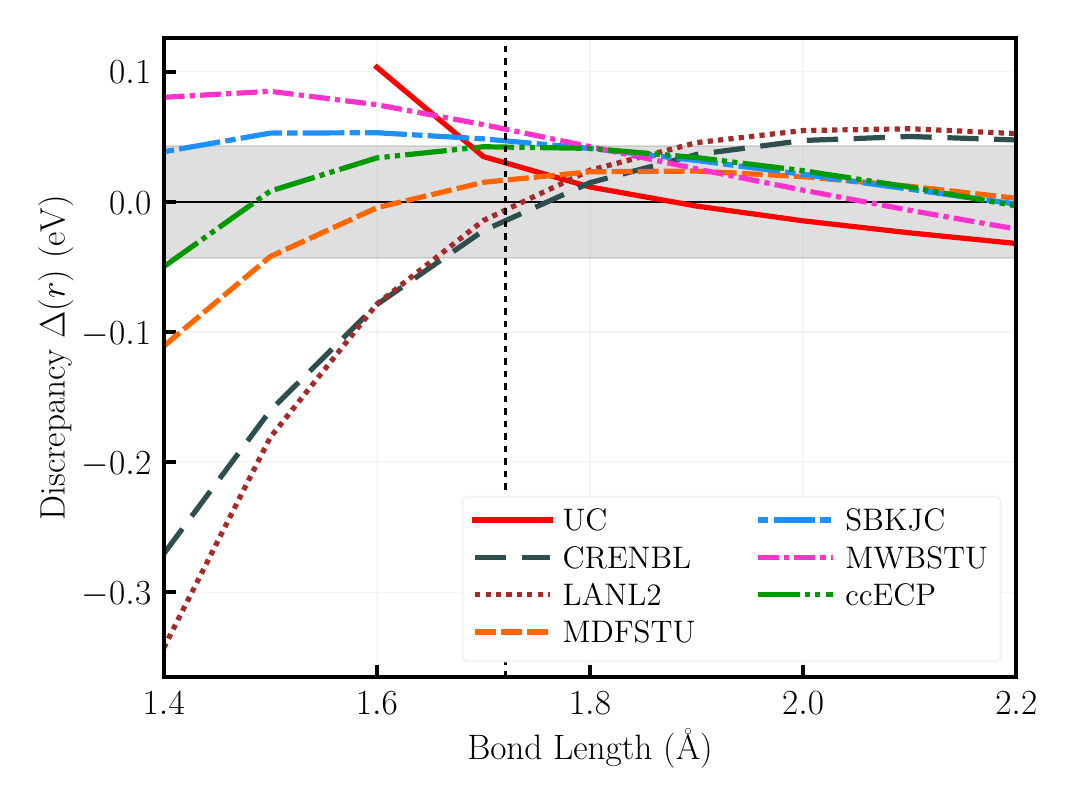}
\caption{ReO binding curve discrepancies}
\label{fig:ReO}
\end{subfigure}
\caption{Binding energy discrepancies for (a) ReH and (b) ReO molecules. The grey-shaded region indicates the chemical accuracy, and the vertical dashed line gives the equilibrium bond length from the AE result.
}
\label{fig:Re_mols}
\end{figure*}

\begin{figure*}[!htbp]
\centering
\begin{subfigure}{0.5\textwidth}
\includegraphics[width=\textwidth]{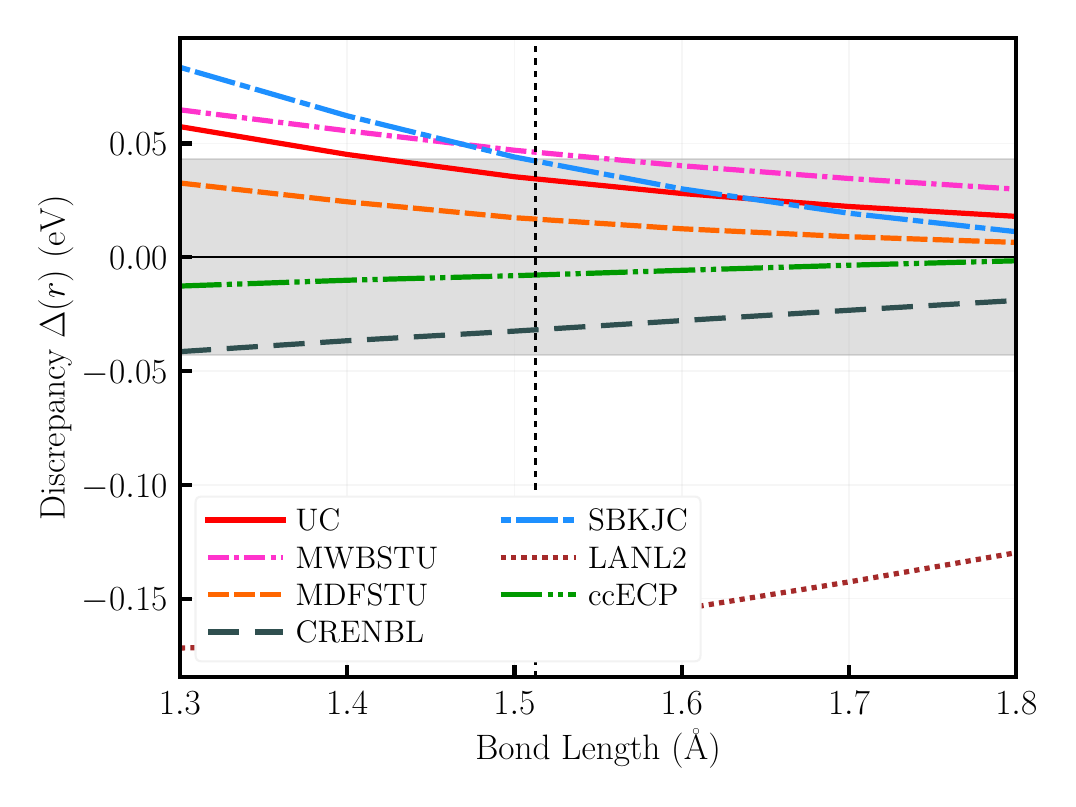}
\caption{PtH binding curve discrepancies}
\label{fig:PtH}
\end{subfigure}%
\begin{subfigure}{0.5\textwidth}
\includegraphics[width=\textwidth]{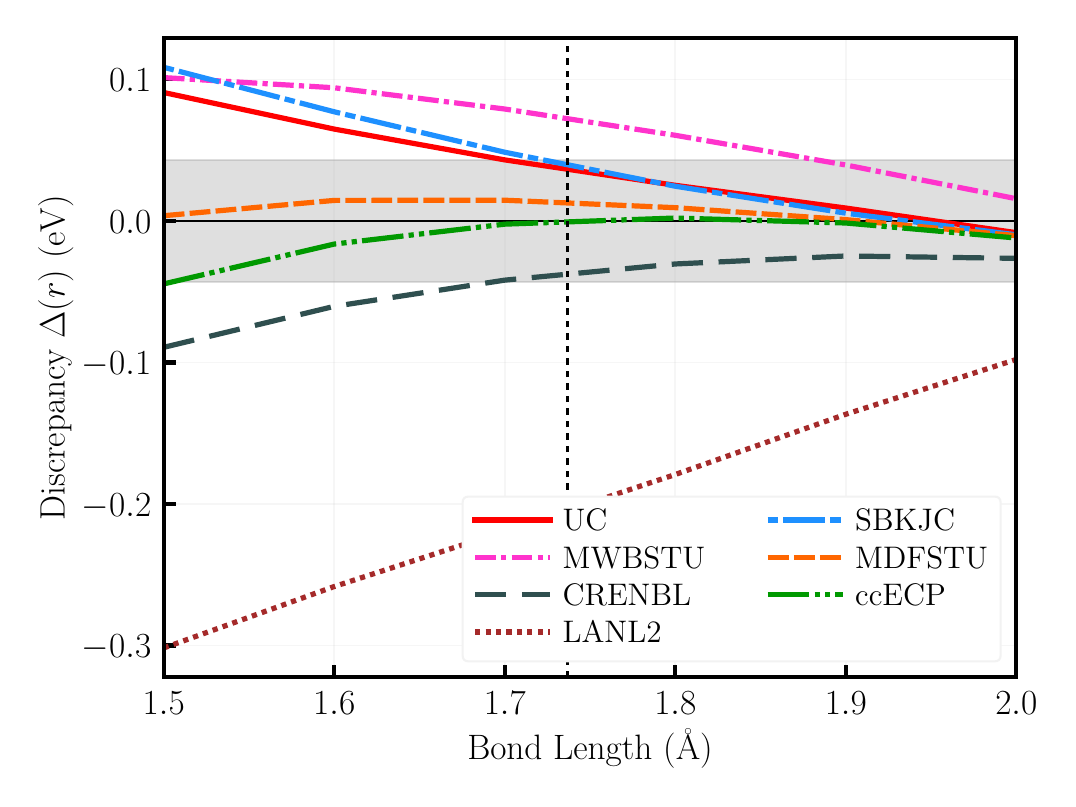}
\caption{PtO binding curve discrepancies}
\label{fig:PtO}
\end{subfigure}
\caption{Binding energy discrepancies for (a) PtH and (b) PtO molecules. The grey-shaded region indicates the chemical accuracy, and the vertical dashed line gives the equilibrium bond length from the AE result.}
\label{fig:Pt_mols}
\end{figure*}

For the molecular calculation, similar to the $4d$ series, we also observed overall consistency and  quality improvements shown in Figures \ref{fig:Re_mols}, \ref{fig:Ta_mols}, and \ref{fig:Pt_mols}. 
For Re, the ReH and ReO ccECP binding curves are the only pair of curves that remain within the chemical accuracy across the range of bond length, with both ReH and ReO exhibiting noticeable improvements compared with other ECPs. 
It is worthwhile to mention that in ReO for UC, we exclude the molecular binding discrepancy curve near dissociation bond length due to the difficult convergence in the coupled cluster calculations.
The uncorrelated core shows notable underbinding at short bond lengths.
We conjecture that this is caused by the rigidity of $4f$ orbitals with long tails that reach the bonding region.
This suggests that the correlations related to this subspace need to be captured, at least indirectly. 
For Ta, the TaH binding curve is on par with MDFSTU and MWBSTU, consistently within the chemical accuracy for the entire bond length range. 
For TaO, ccECP binding curve shows better accuracy, especially at the equilibrium bond length, while alleviating the overbinding issue in LANL2, MDFSTU, and CRENBL. 
For Pt, it is evident that our ccECP, alongside MDFSTU, outperforms other candidates across the entire bond length range.


It is important to note that throughout the optimization of the $5d$ elements, we imposed additional restrictions on the range of all exponents ($\alpha_{lk}$) to prevent high plane wave cutoffs in codes with periodic boundary conditions. Consequently, some compromises observed in the results stem from these additional restrictions made in favor of wider applicability in plane wave applications.

In conclusion, our ccECPs for the selected 5$d$ elements, Ta, Re, and Pt, exhibit systematic improvements in both atomic spectra accuracy and molecular binding curves.


\begin{table}[!htbp]
\centering
\caption{Mean absolute deviations of molecular binding parameters for various core approximations with respect to AE data for $5d$ selected heavy-element groups, Re, Ta, and Rh molecules. All parameters were obtained using Morse potential fit. The parameters shown are dissociation energy $D_e$, equilibrium bond length $r_e$, vibrational frequency $\omega_e$ and binding energy discrepancy at dissociation bond length $D_{diss}$.}
\label{morse:additional}
\begin{tabular}{l|rrrrrrrrrr}
\hline\hline
{} & $D_e$(eV) & $r_e$(\AA) & $\omega_e$(cm$^{-1}$) & $D_{diss}$(eV) \\
\hline
UC     &  0.02(1) &  0.008(5) &                21(34) &       0.30(12) \\
CRENBL &  0.04(2) &  0.005(5) &                10(37) &       0.14(11) \\
LANL2  &  0.11(2) &  0.008(5) &                28(38) &       0.25(11) \\
MDFSTU &  0.03(2) &  0.005(5) &                13(35) &       0.09(11) \\
MWBSTU &  0.04(2) &  0.009(5) &                29(34) &       0.08(11) \\
SBKJC  &  0.07(2) &  0.007(5) &                35(36) &       0.10(12) \\
ccECP  &  0.02(2) &  0.003(5) &                14(34) &       0.06(11) \\
\hline\hline
\end{tabular}
\end{table}
\subsection{Selected $4f$ elements}
The elements in this section are characterized by a unique core configuration, which represents one of the novel aspects of this work.
The core comprises 46 electrons, separating thus the $n=4$ shell into subshells.
Specifically, the core is given as [[Kr]$4d^{10}$], with the $4f$ subshell included in the valence space. 
This unorthodox core choice has been less explored in previous ECP constructions
since the core is often defined by a closed shell of principal quantum number $n$.
For instance, the core for $3d$ transition metals includes electrons up to $n=2$ ([Ne]-core), while for the $4d$ transition metals it is $n=3$ ([Ar]$3d^{10}$).

Generating ECPs for these elements required a modified approach due to the chosen core structure. 
We first fitted eigenvalues for each one-particle channel to establish reliable initial guess parameters.
A similar procedure of developing initial guess parameters has been applied in our previous work when constructing soft ccECPs for the late $3d$ transition metals \cite{Kincaid2022}, and an analogous initialization was originally proposed by Dolg et al. \cite{Dolg1987}.
We refer the reader to Section \ref{ECP_init_5f} for details of ECP parameter initialization descriptions.
The subsequent optimization procedure follows a similar scheme to our previous work, where the energy gaps for various states served as the objective function for optimization, iterated until the desired accuracy for the atomic spectrum was achieved.
Thereafter, we performed the molecular transferability tests and made adjustments to obtain the best possible balance between atomic spectrum accuracy, molecular binding curve accuracy, and plane wave cutoffs as an additional criterium. 
Compared to previous sets of heavier elements \cite{Wang2022},  we used a smaller number of highly-ionized states due to the substantial difficulties they have caused in the optimization process.



Despite the additional constraints applied to exponents ($\alpha_{lk}$) for a desirably small cutoff in plane wave applications, our ccECP atomic results for Tb display comparable accuracy with the uncorrelated core results and show a significant improvement over SBKJC and also DFT-PBE (which was generated using Troullier-Martins scheme \cite{Troullier1991} and PBE functional \cite{Perdew1996}) in MAD, LMAD, and WMAD characteristics.
In the case of Gd, the atomic results are again comparable to the UC results.
As also observed for Tb, the UC atomic spectra accuracy is surprisingly high and very challenging to reach in our optimizations. This may indicate that our ccECP for Gd is probably approaching the fidelity limit of the presented construction.

\begin{figure*}[!htbp]
\centering
\begin{subfigure}{0.5\textwidth}
\includegraphics[width=\textwidth]{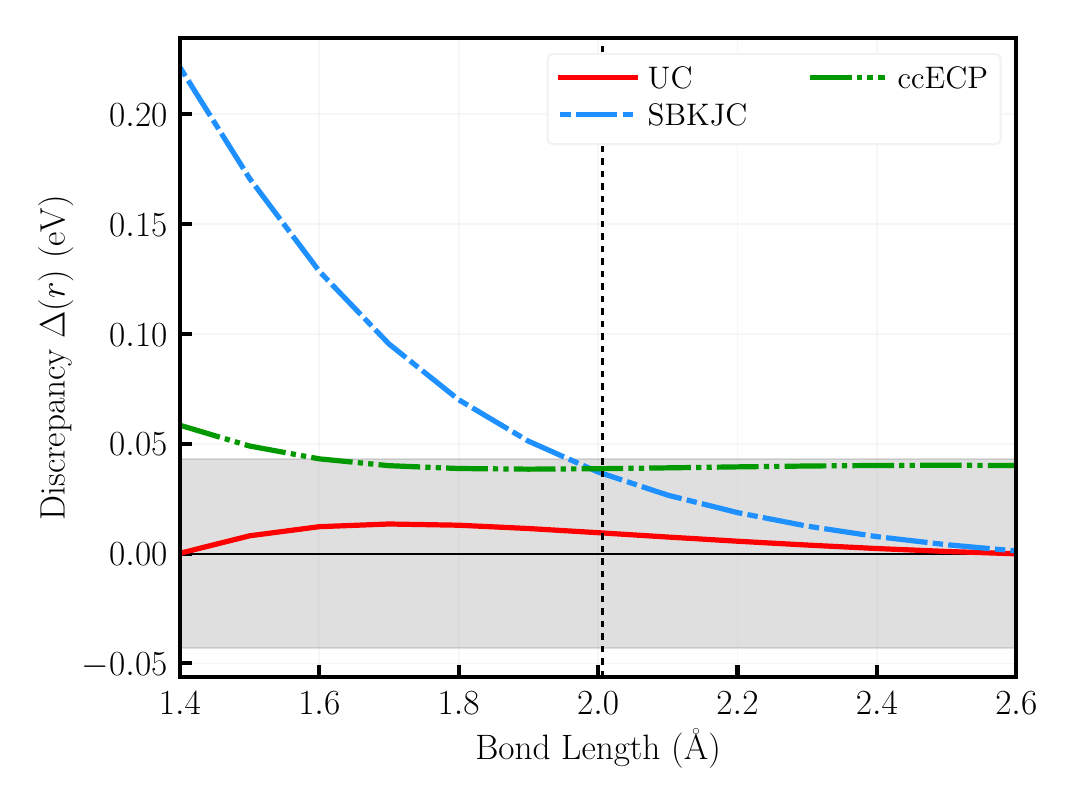}
\caption{GdH$_3$ binding curve discrepancies}
\label{fig:GdH}
\end{subfigure}%
\begin{subfigure}{0.5\textwidth}
\includegraphics[width=\textwidth]{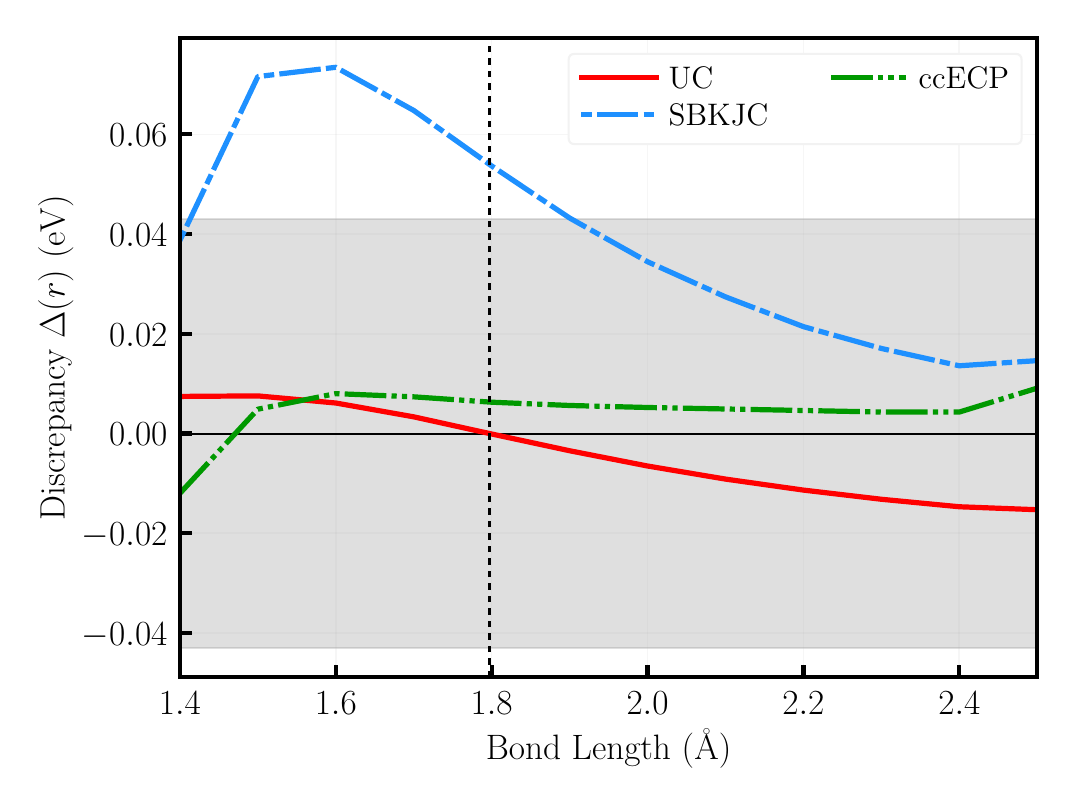}
\caption{GdO binding curve discrepancies}
\label{fig:GdO}
\end{subfigure}
\caption{Binding energy discrepancies for (a) GdH$_3$ and (b) GdO molecules. The grey-shaded region indicates the chemical accuracy, and the vertical dashed line gives the equilibrium bond length from the AE result.
}
\label{fig:Gd_mols}
\end{figure*}

\begin{figure*}[!htbp]
\centering
\begin{subfigure}{0.5\textwidth}
\includegraphics[width=\textwidth]{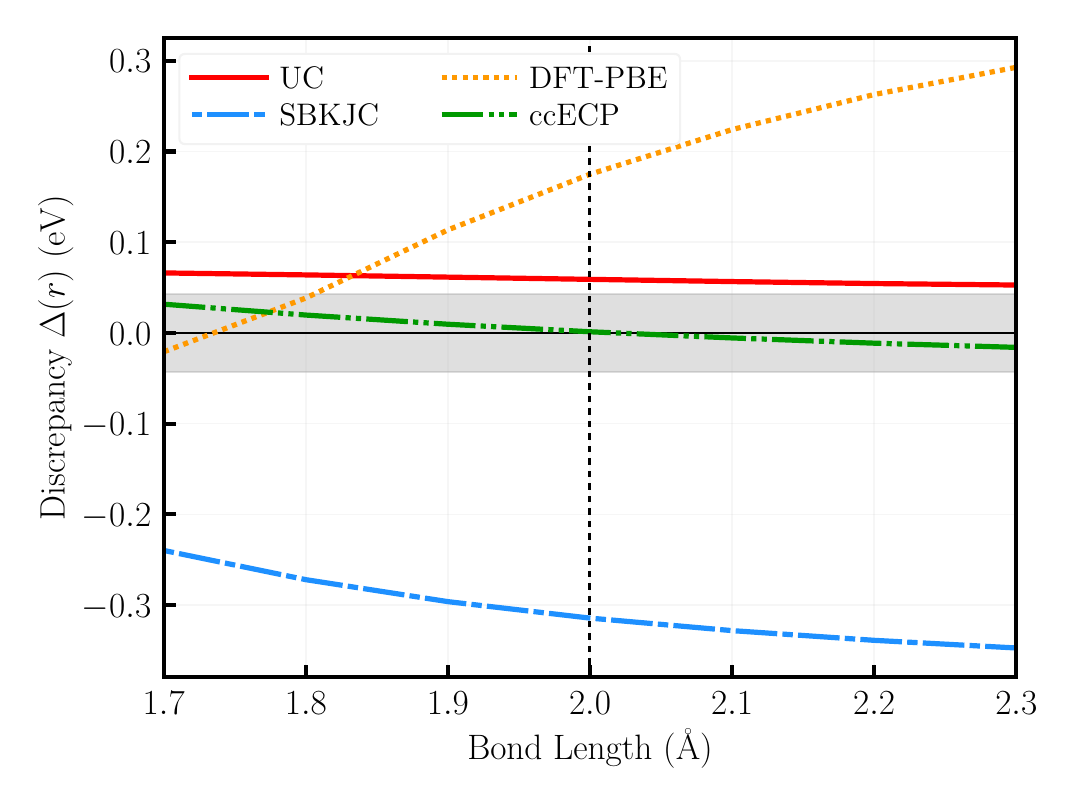}
\caption{TbH$_3$ binding curve discrepancies}
\label{fig:TbH}
\end{subfigure}%
\begin{subfigure}{0.5\textwidth}
\includegraphics[width=\textwidth]{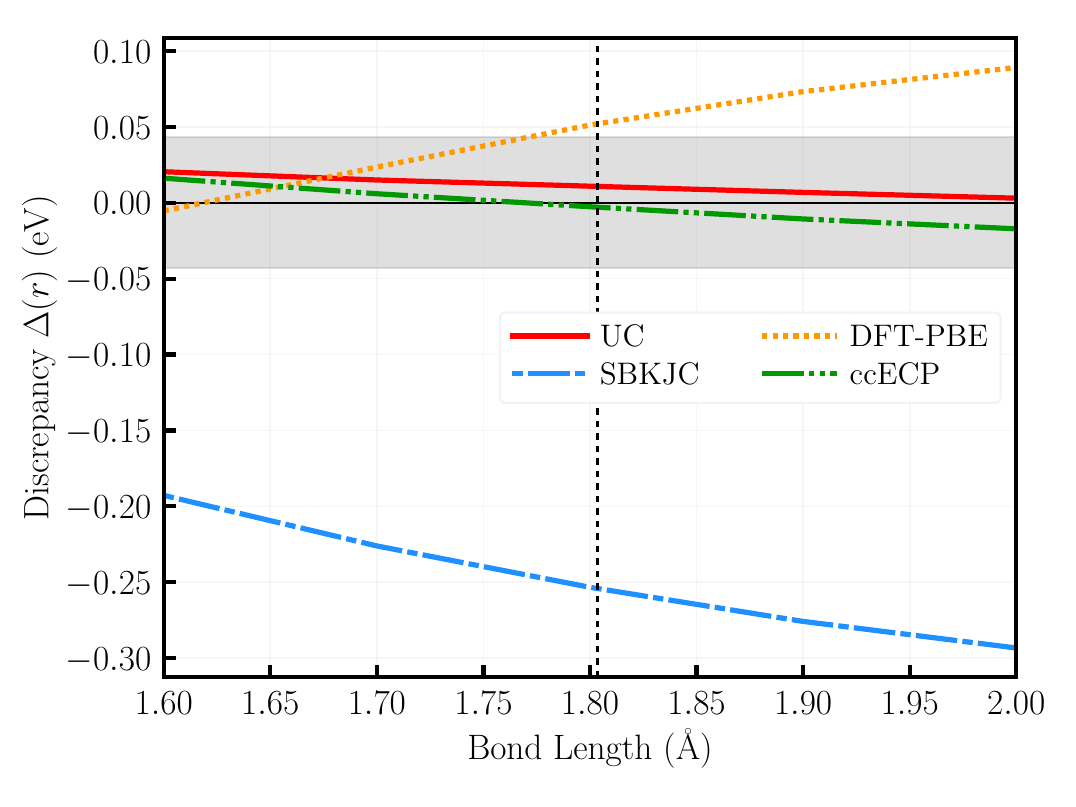}
\caption{TbO binding curve discrepancies}
\label{fig:TbO}
\end{subfigure}
\caption{Binding energy discrepancies for (a) TbH$_3$ and (b) TbO molecules. The grey-shaded region indicates the chemical accuracy, and the vertical dashed line gives the equilibrium bond length from the AE result.
DFT-PBE is an ECP based on the Troullier-Martins scheme \cite{Troullier1991}, and PBE functional \cite{Perdew1996}.
}
\label{fig:Tb_mols}
\end{figure*}

For transferability tests of rare-earth elements (R), we focused on RH$_3$ and RO molecules, instead of regular monohydride and monoxide molecules since the R$^{2+}$ and R$^{3+}$ are more relevant for wider applications, such as in bulk materials.
The results are summarized in Figures \ref{fig:Gd_mols} and \ref{fig:Tb_mols}.
For both TbH$_3$ and TbO, our ccECP is clearly surpassing the legacy ECPs in quality and is comparable to the UC treatment.
For most of the bond lengths, GdH$_3$ ccECP binding curves are within the chemical accuracy with minor compromises compared to UC calculations.
Overall, our tests demonstrate successful, high-accuracy ccECP generation viability for rare-earth elements, utilizing Gd and Tb and the core choice that hasn't been systematically explored before.
In addition, we find that the accuracy of UC is unexpectedly high in Gd and Tb, which is at odds with some previous observations such as large core Na-Ar elements where ccECP accuracy was considerably better than UC \cite{Bennett2018}.



We note that further studies are still needed to better understand Gd, Tb, and rare-earth elements in general.
For example, just identifying the ground state of an isolated Tb atom proved to be non-trivial.
Spectroscopy experiments indicate that the correct ground state is [Xe]$4f^9 6s^2$ ($J=\frac{15}{2}$) and the lowest excited state is [Xe]$4f^8 5d 6s^2$ ($J=\frac{13}{2}$) with a $\approx 0.035$~eV gap \cite{A.Kramida2022}.
Scalar relativistic AE HF calculations obtain a gap of $\approx -3.25$~eV (negative, wrong ground state).
Account of spin-orbit effects at the COSCI level slightly improves the gap to $\approx -2.93$~eV.
Finally, scalar relativistic CCSD(T) obtains a $\approx -0.58$~eV gap.
Clearly, an accurate description of both spin-orbit effects and electron correlations is required to identify the ground state of Tb correctly.
Therefore, the optimized rare-earth ccECPs open possibilities for further studies of spin-orbit effects, electron correlations, and their mutual interplay with more sophisticated approaches such as multireference CI and QMC methods.

\section{Conclusions}
We present a new set of correlation-consistent ECPs
for selected elements Y, Zr, Nb, Rh, Ta, Re, Pt, and lanthanides, Tb, Gd. 
This choice has been motivated by needs from areas such as catalysis and materials research, where these elements play a prominent role. 
We encountered several complications that made construction much more complex than in our previous works on 5$d$ transition metals\cite{Wang2022}.
This required a careful balance between competing criteria such as fundamental accuracy, transferability in various chemical settings, efficiency in subsequent fully correlated approaches, and broad usefulness.

One of the key  difficulties observed for $f$-elements is the near-degeneracy of $4f$ levels with $s$, $p$, and $d$ channel states.
In CCSD(T) calculations, this often results in severe convergence problems in both ECP and all electron calculations.
Additionally, different choices of basis sets can also contribute to the ill-convergence behavior, with the $f$-elements displaying particular sensitivity to the choice of basis sets \cite{Bauschlicher2022}.
During the development, extensive efforts have been devoted to resolving such convergence and state-targeting instability issues. 

We also observe that the usefulness of atomic spectra as the guiding principle of ECP constructions becomes less impactful on final quality than in previous works \cite{Bennett2017, Bennett2018, Annaberdiyev2018, Wang2019, Wang2022}. 
Much more significant are tests on systems with bonds due to the fact that hybridization becomes very complex and involves a large number of one-particle atomic channels due to energetic closeness of $4f, 5d, 6s$, and $6p$ levels.
The relaxation effects and overall restructuring in these channels are substantial and difficult to predict, so the  prominence of probing bonded environments grows.
Indeed, this is reflected in our objective functions and makes a significant difference compared to  previous constructions.
This is also a significant departure from lighter elements, making the optimization more complex and demanding.
The tiny biases of highly ionized states with $f$-occupations are negligible since their contributions in typical bonding situations become marginal, definitely well below other systematic errors involved.

Clearly, recreating the effect of all-electron atoms with a full account of relativity, possibly QED effects, and a full account of correlation poses a significant challenge that is further complicated by effects such as core polarization and relaxation.
One possible remedy would be to include core polarization and relaxation terms that have been occasionally used before. 
Unfortunately, this goes against the simplicity of use since almost all of the existing mainstream condensed matter codes (\textsc{Quantum Espresso}, \textsc{VASP}) and many quantum chemistry basis set packages (\textsc{PySCF}\cite{Sun2017}, \textsc{Gaussian}\cite{Frisch2016}, \textsc{Turbomole}\cite{Turbomole2017}, \textsc{Crystal}\cite{Erba0, Dovesi2020}) do not have such terms implemented (as of this writing).
We note that their inclusion in QMC codes is straightforward (a few lines of code), even in a posteriori manner, as was demonstrated a long time ago \cite{Shirley1991}.
At this point, we were interested in presenting ccECPs in the same form as used  for the lighter elements in order to keep the consistency and systematicity of our constructions.
This  leaves explorations of core relaxation and polarization effects for future work. 
Similarly, we plan to use the constructed effective Hamiltonians in QMC future studies so as to expand the testing and benchmarking in a variety of chemical applications.

Overall, our work provides a new generation of ECPs for selected set of heavy elements, pioneers new core-valence partitioning for $4f$ open-shell atoms, and opens possibilities for valence-only calculations with significantly better account of accuracy and transparency of systematic errors involved.

\section{Supplementary Material}

Additional information about ccECPs can be found in the Supplementary Material.
Therein, AREP results for calculated atomic spectra for each element and molecular binding curve fit parameters are provided.
The atomic spectra results include AE spectra for each element and the corresponding discrepancies of various core approximations.
The molecular fit parameters for hydrides, trihyrides, and oxides are provided.
The ccECPs in semilocal form, Kleinman-Bylander projected forms, as well as optimized Gaussian valence basis sets in various input formats (\textsc{Molpro}, \textsc{GAMESS}, \textsc{NWChem}) can be found on the website \cite{pseudopotentiallibrary}.

Concerning the SO results, the detailed atomic gaps at the COSCI (complete open-shell configuration interaction) level used for optimizing spin-orbit terms are provided.

Input and output data generated and related to this work will be published in Material Data Facility \cite{Blaiszik2016, Blaiszik2019} and can be found in Ref.


\section{Conflict of Interest}
The authors have no conflicts to disclose.

\begin{acknowledgements}

We thank Paul R. C. Kent for reading the manuscript and providing helpful suggestions.

This work has been supported by the U.S. Department of Energy, Office of Science, Basic Energy Sciences, Materials Sciences and Engineering Division, as part of the Computational Materials Sciences Program and Center for Predictive Simulation of Functional Materials.

This research used resources of the National Energy Research Scientific Computing Center (NERSC), a U.S. Department of Energy Office of Science User Facility operated under Contract No. DE-AC02-05CH11231.

An award of computer time was provided by the Innovative and Novel Computational Impact on Theory and Experiment (INCITE) program.
This research used resources of the Oak Ridge Leadership Computing Facility, which is a DOE Office of Science User Facility supported under Contract No. DE-AC05-00OR22725.


This paper describes objective technical results and analysis. Any subjective views or opinions that might be expressed in the paper do not necessarily represent the views of the U.S. Department of Energy or the United States Government.

Notice:  This manuscript has been authored by UT-Battelle, LLC, under contract DE-AC05-00OR22725 with the US Department of Energy (DOE). The US government retains and the publisher, by accepting the article for publication, acknowledges that the US government retains a nonexclusive, paid-up, irrevocable, worldwide license to publish or reproduce the published form of this manuscript, or allow others to do so, for US government purposes. DOE will provide public access to these results of federally sponsored research in accordance with the DOE Public Access Plan (http://energy.gov/downloads/doe-public-access-plan).

\end{acknowledgements}

\bibliographystyle{unsrt}
\bibliography{version1.bib}

\end{document}